\documentclass[twocolumn,showpacs,preprintnumbers,amsmath,amssymb,aps,prb,superscriptaddress]{revtex4-1}

\pdfoutput=1
\usepackage[pdftex]{graphicx}

\usepackage{dcolumn}
\usepackage{bm}

\usepackage{color}
\usepackage{amstext}
\usepackage[utf8]{inputenc}

\newcommand{\e}[1] {\text{e}^{#1}}
\newcommand{\I}[0] {\text{i}}
\newcommand{\D}[0] {\text{d}}

\newcommand{\meanline}[1] {\overline{#1}}
\newcommand{\abs}[1] {\left|#1\right|}

\begin{document}

\title{Interacting spinless fermions in a diamond chain}

\author{A. A. Lopes}
\affiliation{Institute for Molecules and Materials, Radboud University Nijmegen, \\
Heyendaalseweg 135, 6525 AJ Nijmegen, The Netherlands}

\author{R. G. Dias}
\affiliation{Department of Physics, I3N, University of Aveiro,\\
Campus de Santiago, Portugal}

\date{\today}

\begin{abstract}
We study spinless fermions in a flux threaded AB$_2$ chain taking into account nearest-neighbor Coulomb interactions. The exact diagonalization of the spinless AB$_2$ chain is presented in the limiting cases of infinite or zero nearest-neighbor Coulomb repulsion for any filling. Without interactions, the AB$_2$ chain has a flat band even in the presence of magnetic flux. We show that the respective localized states can be written in the most compact form as standing waves in one or two consecutive plaquettes. We show that this result is easily generalized to other frustrated lattices such as the Lieb lattice. 
A restricted Hartree-Fock study of the $V/t$ versus filling phase diagram of the AB$_2$ chain has also been carried out. The validity of the mean-field approach is discussed taking into account the exact results in the case of infinite repulsion. The ground-state energy as a function of filling and interaction $V$ is determined using the mean-field approach and exactly for infinite or zero $V$. 
In the strong-coupling limit, two kinds of localized states occur: one-particle localized states due to geometry and two-particle localized states due to interaction and geometry. These localized fermions create open boundary regions for itinerant carriers.  At filling $\rho=2/9$ and in order to avoid the existence of itinerant fermions  with positive kinetic energy, phase separation occurs between a high-density phase ($\rho=2/3$) and a low-density phase ($\rho=2/9$) leading to a  metal-insulator transition. The ground-state energy reflects such phase separation by becoming linear on filling above $2/9$.
We argue that for filling near or larger than 2/9, the spectrum of the t-V AB$_2$ chain can be viewed as a mix of the spectra of Luttinger liquids (LL) with different fillings,  boundary conditions, and  LL velocities.

\end{abstract}
\pacs{71.10.Fd, 71.10.Hf, 71.10.Pm, 75.10.Pq}

\maketitle

\section{Introduction}

The field of itinerant geometrically frustrated electronic systems has attracted considerable interest in the last two decades \cite{Gul'acsi2007,Kikuchi2005,Macedo1995,Montenegro-Filho2006,Vidal1998,Tamura2002,Tasaki1998,*Tasaki1998a,*Tasaki1998b,
*Tasaki1995,*Tasaki1994,*Tasaki1992,Mielke1999,*Mielke1992,*Mielke1991,*Mielke1991a,*Mielke1993,Derzhko2005,
Derzhko2004,Derzhko2010,Tanaka2007,Duan2001,Gul'acsi2005,Gul'acsi2003,
Richter2004a,Richter2004,Rule2008,Schulenburg2002,
Vidal1998,Wu2007,Wu2008,Zhitomirsky2007}.
A simple example of a geometrically frustrated lattice is the AB$_2$ chain, also designated by  diamond chain or bipartite lozenge chain \cite{Vidal1998,Montenegro-Filho2006,Macedo1995,Gul'acsi2007,Kikuchi2005}.
This is a  quasi-one-dimensional system consisting of an one-dimensional array of quantum rings.  
Such system can be used to model ML$_2$ (metal-ligand) chains \cite{Silvestre1985} and azurite\cite{Kikuchi2005} and can be generalized to molecular systems displaying similar topology \cite{Zheng2007}. Also, given nowadays nanofabrication techniques, such as electron beam lithography, a diamond chain-like system can, in principle, be built from scratch \cite{Tamura2002}.
A particular feature of the  band structure of these frustrated systems is the presence of one or several flat bands in the one-particle energy dispersion, which reflect the existence of localized eigenstates of the geometrically frustrated tight-binding Hamiltonian \cite{Mielke1999,Mielke1992,Mielke1991,Mielke1991a,Mielke1993,Tasaki1998,Tasaki1998a,Tasaki1998b,
Tasaki1995,Tasaki1994,Tasaki1992,Vidal1998,Montenegro-Filho2006,Macedo1995, Derzhko2005,
Derzhko2004,Derzhko2010,Gul'acsi2005,Gul'acsi2003,Gul'acsi2007,Duan2001,Kikuchi2005,Tamura2002,Tanaka2007,Vidal1998}.

The ground states of the Hubbard model for an AB$_2$ geometry are well studied \cite{Vidal1998,Montenegro-Filho2006,Macedo1995, Gul'acsi2007,Kikuchi2005} as well as  for other frustrated  lattices such as the sawtooth chain \cite{Derzhko2010}, the Kagome chain \cite{Derzhko2010}, etc. Some of these frustrated lattices fall onto the category of the cell construction lattices proposed by Tasaki \cite{Tasaki1998,Tasaki1998a,Tasaki1998b,Tasaki1995,Tasaki1994,Tasaki1992} or the category of  line graphs lattices proposed by Mielke  \cite{Mielke1999,Mielke1992,Mielke1991,Mielke1991a}. 
The approach followed in some of these studies relies in the fact that the Hubbard interaction is a positive semidefinite operator and is limited to cases where the lowest band is a flat band or the chemical potential is fixed at the flat-band energy \cite{Gul'acsi2007}.
The frustrated systems have  usually been studied in the absence of flux \cite{Montenegro-Filho2006,Macedo1995,Duan2001}.
In the case of AB$_2$ Hubbard chain, the flux dependence of the ground-state energy has been studied but again only for chemical potential fixed at the flat band energy  \cite{Gul'acsi2007}. 
Many different ground states are possible in the AB$_2$ Hubbard chain, leading to a great variety of properties such as flat-band ferromagnetism or  half-metallic conduction, depending on the values of filling, interaction or magnetic field \cite{Gul'acsi2007}.

The case of spinless fermions in a AB$_2$ lattice, taking into account   nearest-neighbor Coulomb interaction, is simpler than the Hubbard model due to the absence of spin degrees of freedom.  We will designate the respective Hamiltonian for such system by t-V AB$_2$ Hamiltonian.
The t-V  model, in its strictly one-dimensional version (1D), can be mapped into the anisotropic Heisenberg model (more precisely, the XXZ or Heisenberg-Ising model) by the Jordan-Wigner transformation \cite{Jordan1928}, whose Bethe ansatz solution has long been known \cite{Baxter1973,*Baxter1973a,*Baxter1973b,*Bethe1931}. For quasi-1D models such as the one discussed in this manuscript, a Bethe ansatz solution is not possible. However, a Jordan-Wigner transformation into the XXZ AB$_2$ chain should be possible using  the extension of the  Jordan-Wigner transformation to two dimensions which has been discussed by several authors \cite{Fradkin1989,Wang1992,*Wang1992a,*Wang1991,Ambjoern1989}. In the case of a square lattice, this transformation requires the introduction of a gauge field which in contrast to the one-dimensional case affects the energy spectrum \cite{Wang1992,*Wang1992a,*Wang1991}. In particular, the strong-coupling limit of the repulsive t-V model under a Jordan-Wigner transformation is mapped into the strongly anisotropic antiferromagnetic Heisenberg model.
During the last years, many studies of the antiferromagnetic Heisenberg model in geometrically frustrated lattices  have been carried out \cite{Derzhko2005,Derzhko2004,Duan2001,Kikuchi2005,Macedo1995,Moessner2006,Richter2004,Richter2004a,Rule2008,Schulenburg2002,Zhitomirsky2007}.
Under high magnetic fields but below the saturation field, the ground states of these frustrated magnetic systems consist of localized and independent magnons created in a ferromagnetic background. As a consequence of these localized magnons, quantized plateaus have been 
found in the respective magnetization curves \cite{Schulenburg2002}.

In this manuscript, the exact diagonalization of the spinless AB$_2$ chain is presented in the limiting cases of infinite or zero nearest-neighbor Coulomb interaction for any filling and in the presence of magnetic flux. Without interactions, the AB$_2$ chain has a flat band even in the presence of magnetic flux.
A simple construction of the localized states that generate the flat bands both in the presence and absence of flux is presented and generalized  to  arrays of AB$_n$ quantum rings. The flat band generates a plateau in the ground-state energy as a function of filling (for fillings between 1/3 and 2/3).
A restricted Hartree-Fock  study of the $V/t$ versus filling  phase diagram  has  been carried out. 
For finite $V$, the mean-field ground-state energy increases in relation to the independent fermions ground-state energy, but remains negative for fillings lower than 2/3. For filling larger than 2/3, it becomes approximately linear reflecting the existence of nearest-neighbor occupied sites. 
In the mean-field phase diagram, a uniform density phase is found at low filling. For filling larger than 1/3, one of the mean-field solutions for the  density difference between A sites and B and C pairs of sites disappears (due to  Pauli's exclusion principle).
The validity of the mean-field approach is discussed taking into account the exact results in the  limiting cases.

The strong-coupling limit is particularly interesting due to the presence of two kinds of localized states: one-particle localized states due to geometry and two-particle localized states due to interaction and geometry. Localized states may lead to flat bands or mix with itinerant states creating open-boundary regions for itinerant carriers. 
The localized fermions due to geometry that appear in the frustrated AB$_2$ chain have a direct correspondence with the independent localized magnons in the near-saturation magnetically frustrated systems. Note, however, that the equivalent to the two-particle localized fermions in the frustrated AB$_2$ chain has never been mentioned in studies of the near-saturation  antiferromagnetic Heisenberg model in geometrically frustrated lattices, as far as we know.
The zero-temperature phase diagram obtained plotting the ground-state energy as a function of filling displays an interesting quantum critical point at filling $\rho=2/9$ where a metal-insulator transition occurs.
This transition reflects a phase separation between a high density phase ($\rho=2/3$) and a low-density phase ($\rho=2/9$) that occurs at fillings larger than $\rho=2/9$.
It is worthwhile to emphasize that  phase separation is know to occur in the \textit{attractive} t-V chain  \cite{Haldane1980,Nakamura1997} but it does not occur for repulsive nearest-neighbor interactions.
In the t-V AB$_2$ chain, we have shown that phase separation occurs for strong \textit{repulsive} nearest-neighbor  interactions.
The particularity about the filling $\rho=2/9$ is that the reduced effective lattice (infinite nearest-neighbor repulsion effectively reduces the number of available sites for fermions) becomes half-filled at $\rho=2/9$. Phase separation occurs in order to avoid the existence of itinerant fermions with positive kinetic energy.
We also discuss whether  the strong coupling t-V AB$_2$ chain is a Luttinger liquid. 
We argue that while for low filling, the low energy properties of t-V AB$_2$ chain can be described by the spinless Luttinger Hamiltonian,  for filling near or larger than 2/9, the AB$_2$ set of eigenstates and eigenvalues becomes a complex  mix of the sets of eigenstates and eigenvalues of LLs with different sizes, fillings,  boundary conditions, and  LL velocities.

The remaining part of this paper is organized in the following way. In Sec. II, the model is defined.
In Sec. III, the  eigenstates of the AB$_2$ tight-binding model in the presence of magnetic flux are found and the ground-state energy as a function of filling is determined for any flux.  A generalization of the flat-band states  to  AB$_n$ chains or even more complex lattices  is  discussed in Sec. IV.
In Sec. V, the study of the restricted Hartree-Fock phase diagram of the AB$_2$ chain for $V/t$ versus filling is presented. 
In Sec. VI, the strong-coupling limit of the model is addressed. The Luttinger liquid description of the strong coupling t-V AB$_2$ chain is also discussed. In Sec. VII, we show that the results obtained in the previous sections are relevant for the extended Hubbard model in the same geometry. In Sec. VIII, we conclude.

\section{The AB$_2$ chain }

\begin{figure}[t]
	\begin{center}
		$\begin{array}{c@{\hspace{.1in}}c@{\hspace{.1in}}c}
		\includegraphics[width=6cm]{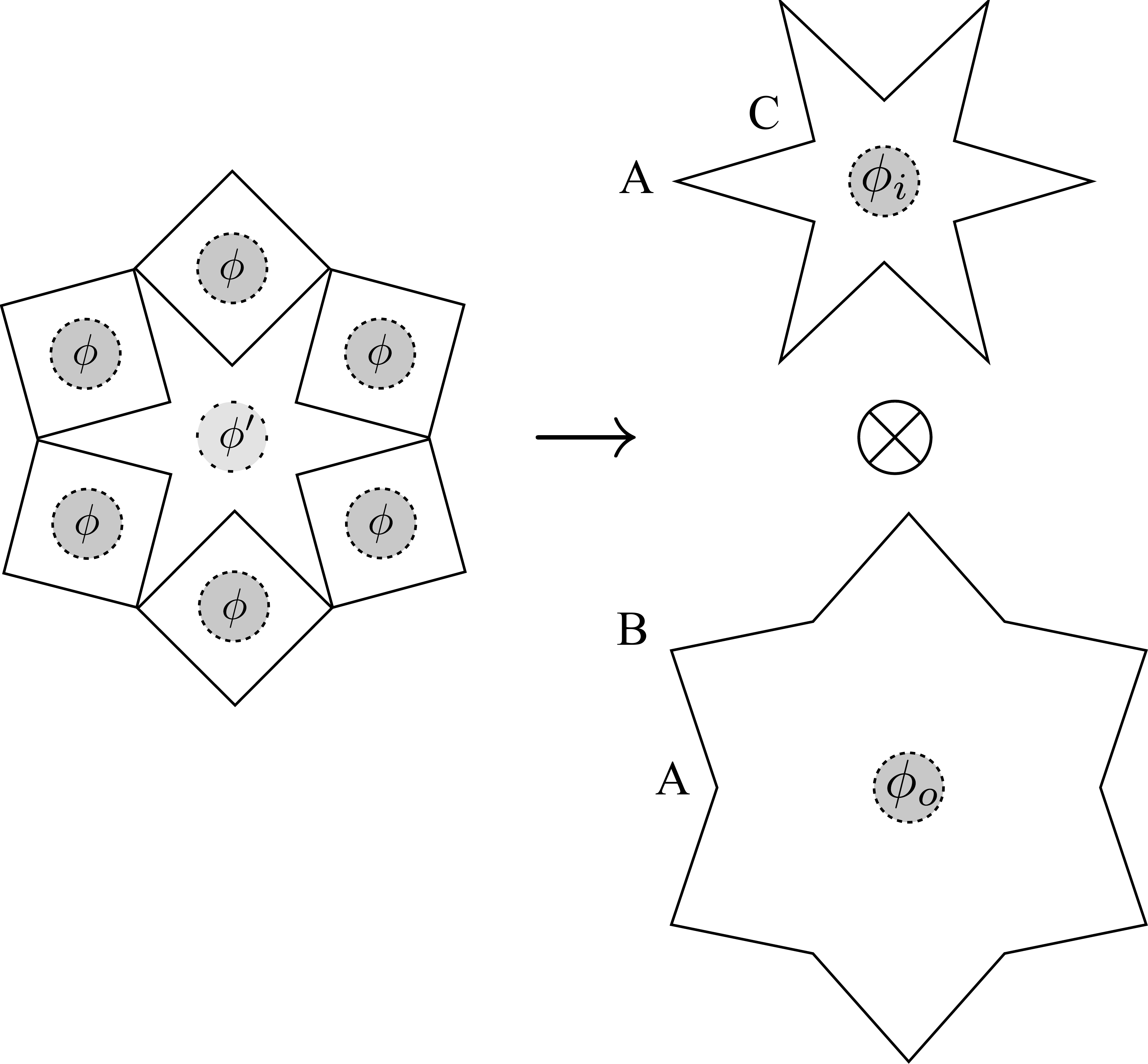}  
		\\		
		(a) \\
		\\
		\includegraphics[width=6cm]{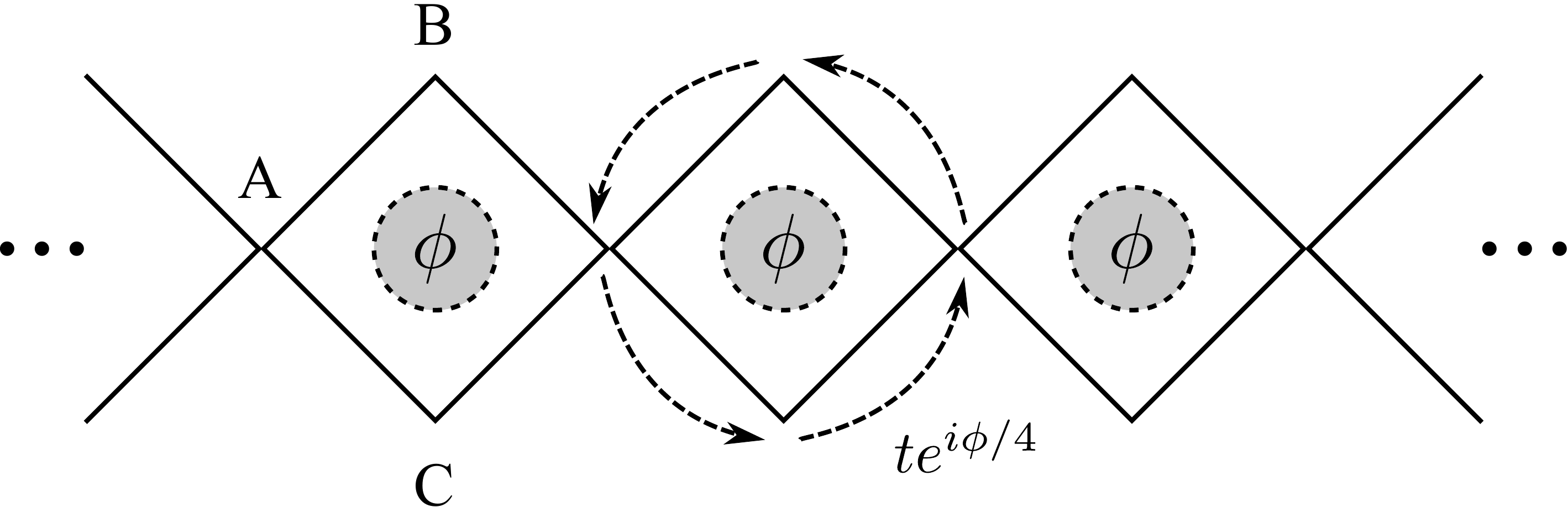}  
		\\
       (b)
		\end{array}$
	\end{center}
    \caption{(a) The AB$_2$ chain consists of an one-dimensional array of diamond rings. Under magnetic field, each diamond ring is threaded by a magnetic flux $\phi$ and the inner star is threaded by a flux $\phi_i$. This system can be pictured as an outer ring with flux $\phi_o$ and an inner ring with a flux $\phi_i$. (b) The diamond chain when $\phi'=0$. }
    \label{fig:star}
\end{figure}

In Fig.~\ref{fig:star}(a), a diamond ring is shown with a magnetic flux $\phi$ threading each diamond plaquette and a magnetic flux $\phi_i$ threading the inner ring. The  inner sites in Fig.~\ref{fig:star}(a) are denoted as C sites and the  outer sites as B sites. Note that there are two ways to close the linear AB$_2$ chain shown in Fig.~\ref{fig:star}(b), either by leaving the B sites or the C sites in the interior of the ring. The two situations are physically equivalent  and therefore we will assume that we have closed our ring so as to leave the C sites as inner sites. The system may be pictured as two rings, an outer one and an inner one, as shown in Fig.~\ref{fig:star}(a), so that  an electron traveling through the outer ring sees an effective flux $\phi_o$ while and electron traveling through the inner ring sees a different effective flux $\phi_i$. 
It will prove itself useful to introduce an auxiliary flux $\phi'$ such that
\begin{equation}
    \begin{split}
        \phi_o = \phi' + N_c \dfrac{\phi}{2}, \\
        \phi_i = \phi' - N_c \dfrac{\phi}{2},
    \end{split}
\end{equation}
where $N_c$ is the number of cells of the diamond ring.
In the case of zero $\phi'$, both effective fluxes still remain non-zero if $\phi$ is nonzero.
In this case, the Peierls phase can be equally distributed in each plaquette, restoring the translational invariance in each ring as shown in Fig.~\ref{fig:star}(b). For $\phi=0$ or $\pi$, the lattice is invariant in the ``flip'' of one plaquette (so that B and C sites exchange places), reflecting a local Z$_2$ symmetry \cite{Doucot2002}.

Considering nearest-neighbor Coulomb interactions and particle hoppings,  the t-V Hamiltonian for an AB$_2$ chain with $N_c$ unit cells (or diamonds) is
\begin{equation}
    H = H_0 + V \sum_j \left(n^A_j + n^A_{j+1}\right)\left( n^B_j + n^C_j \right),
\end{equation}
where $V$ is the value of the interaction and 
\begin{eqnarray}
    H_0 = -t \sum_{j=1}^{N_c} &  & \left[ \e{\I \phi_o/2N_c} ( A_j^\dagger B_j  + B_j^\dagger A_{j+1}) \right. \\
    & + & \left. \e{-\I \phi_i/2N_c} ( C_j^\dagger A_j + A_{j+1}^\dagger C_j ) \right] + \text{H.c.} \nonumber
\end{eqnarray}
Here we have chosen a gauge such that the Peierls phases are equally distributed in the inner ring and in the outer ring.  

\section{Tight-binding limit}

\begin{figure}[t]
    \includegraphics[width=6cm]{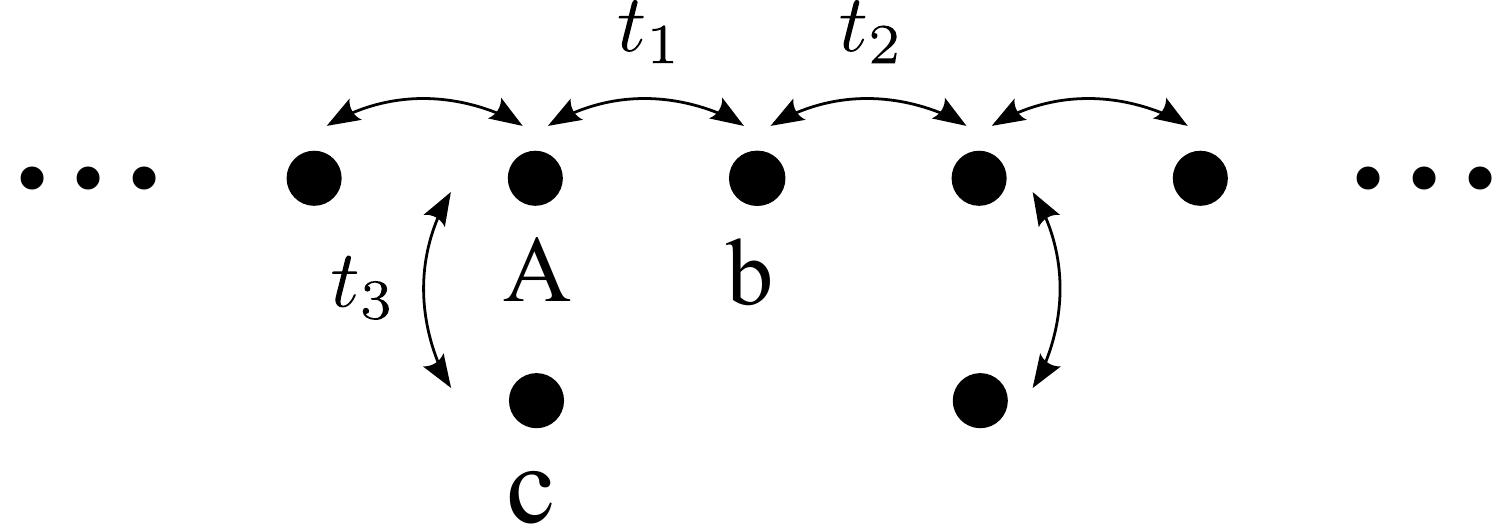}
    \caption{The Hamiltonian of the diamond chain threaded by an arbitrary flux can be mapped into a non-interacting periodic Anderson-like model  with a basis of three sites per unit cell and with  hopping amplitudes  controlled by magnetic flux.}
    \label{fig:Abc}
\end{figure}

\begin{figure}[t]
	\begin{center}
		$\begin{array}{c@{\hspace{.1in}}c@{\hspace{.1in}}c}
		\includegraphics[width=6cm]{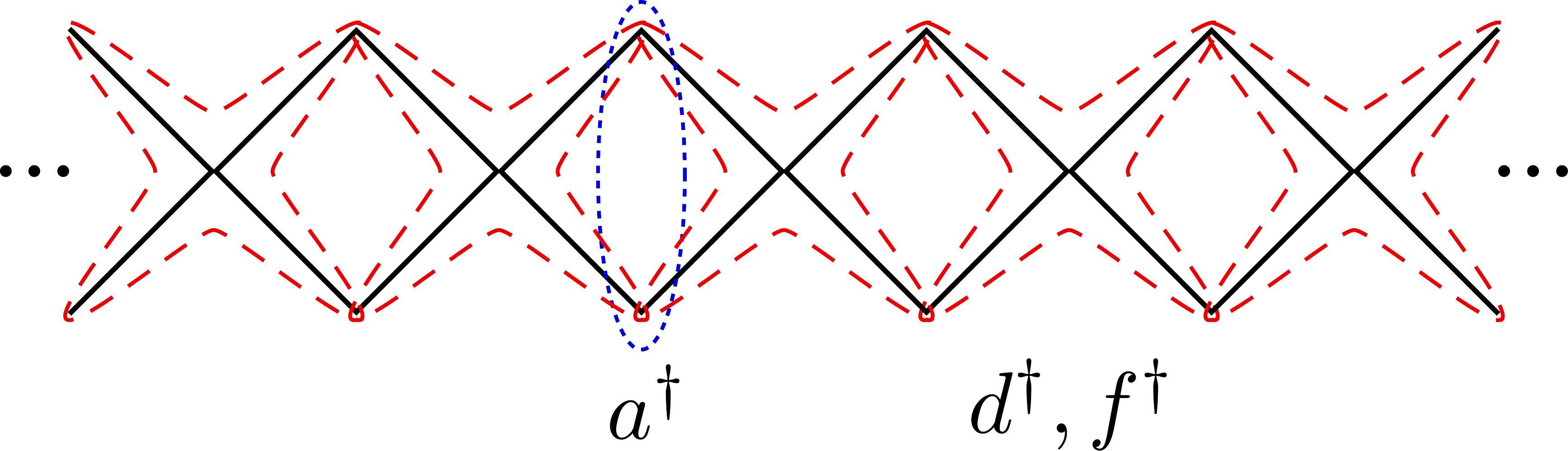}  
		\\		
		(a) \\
		\includegraphics[width=6cm]{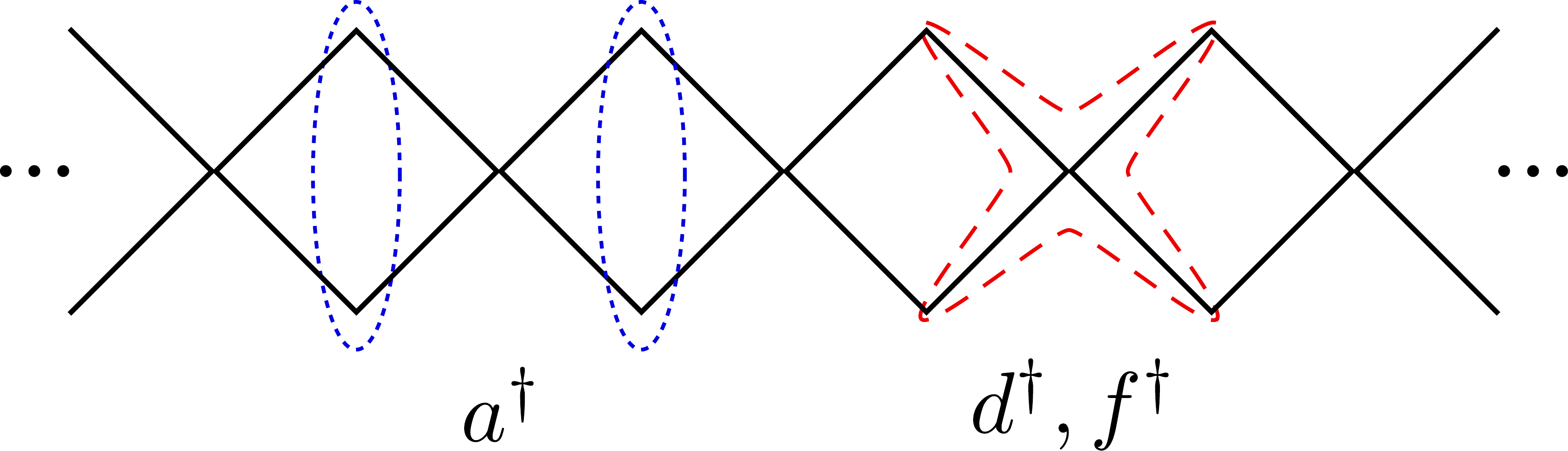}  
		\\
       (b)
		\end{array}$\label{fig:diamondStates0}\label{fig:diamondStatesPi}
	\end{center}
    \caption{Eigenstates of the tight-binding AB$_2$ chain for (a) $\phi = 0$ and (b) $\phi= \pi$. For zero flux, itinerant (red dashed line) as well as  localized eigenstates (blue dotted line) are present. For $\phi= \pi$, all states are localized\cite{Gulacsi2008}.}
    \label{fig:diamondStates}
\end{figure}

\begin{figure}[t]
    \includegraphics[width=8cm]{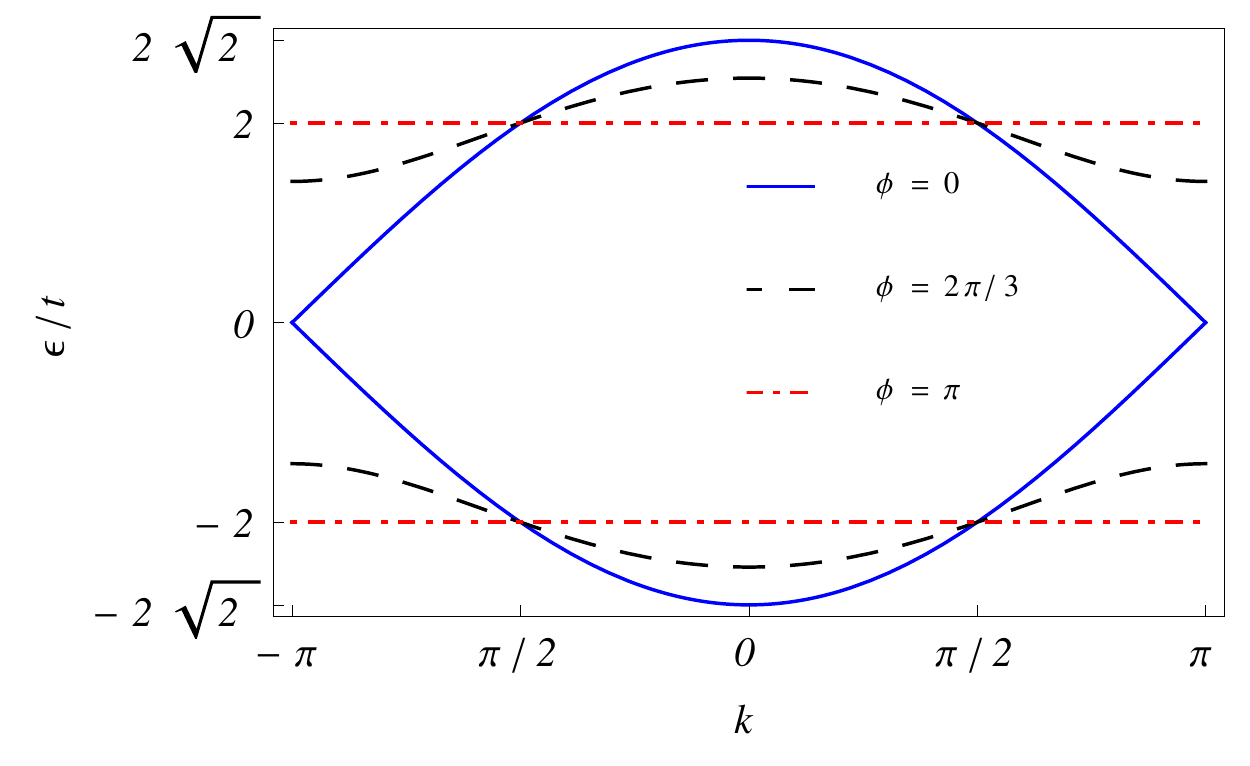}
    \caption{Dispersion relation of the diamond chain for different flux values (we have omitted the flat band $\epsilon_k = 0$ since it is not affected by the flux). For $\phi=\pi$, all bands are flat reflecting the local $Z_2$ symmetry of the Hamiltonian.}
    \label{fig:dRGeneral}
\end{figure}

For simplicity, we will assume a  flux configuration such that $\phi'=0$ in this section. The general case is considered at the end of this section. The  tight-binding Hamiltonian for AB$_2$ chain with $N_c$ cells is
\begin{equation}
  \begin{split}
    H = &-t \sum_j^{N_c} A_j^\dagger \left( \e{ \I \phi/4 } B_j + \e{ -\I \phi/4 } C_j \right) \\
        &+ A_{j+1}^\dagger \left( \e{ -\I \phi/4 } B_j + \e{ \I \phi/4 } C_j \right) + \text{H.c.} 
  \end{split}
\end{equation}
Employing the transformations
\begin{equation}
  \begin{pmatrix}
    b_j^\dagger \\
    c_j^\dagger
  \end{pmatrix}
=\dfrac{1}{\sqrt{2}}
  \begin{pmatrix}
     \e{-\I \phi/4}      & \e{\I \phi / 4} \\
      -\I \e{-\I \phi/4} & \I \e{\I \phi / 4}
  \end{pmatrix}
  \begin{pmatrix}
    B_j^\dagger \\
    C_j^\dagger
  \end{pmatrix},
\end{equation}
the system may be mapped into an Anderson-like model as depicted in Fig. \ref{fig:Abc}. Its Hamiltonian is given by
\begin{equation}
    \begin{split}
    H =  &-t_1 \sum_j b_j^\dagger A_j -t_2 \sum_j b_j^\dagger A_{j+1} \\
	 &-t_3 \sum_j c_j^\dagger A_{j+1} + \text{H.c.},
    \end{split}
\end{equation}
where
\begin{eqnarray}
        t_1 &=& \sqrt{2}t, \\
        t_2 &=& \sqrt{2}\cos(\phi/2)t, \nonumber\\
        t_3 &=& \sqrt{2}\sin(\phi/2)t. \nonumber
\end{eqnarray}
We note that the possibility of mapping a chain Hamiltonian into a periodic Anderson model has been already explored by Gul\'acsi {\it et al} by mapping the triangular-chain Hamiltonian into a periodic Anderson model\cite{Gulacsi2008}. In our case, when the flux is either zero or $\pi$, one has localized eigenstates, which lead to flat bands in the energy dispersion (see Fig.~\ref{fig:dRGeneral}). For zero flux the system becomes a tight-binding ring of A and b sites with independent c sites (see Fig.~\ref{fig:Abc}), having plane-wave eigenstates in the ring and localized eigenstates at c sites, as depicted in Fig. \ref{fig:diamondStates}(a). On the other hand, for $\phi=\pi$, the system behaves as a set of $N_c$ independent  systems with three sites and all the eigenstates are localized as depicted in Fig. \ref{fig:diamondStates}(b)\cite{Gulacsi2008}.
The eigenvalues for an arbitrary value of flux are  given by
\begin{equation}
        \begin{split}
            \epsilon_\text{flat} &= 0, \\
            \epsilon_{\pm} &= \pm 2t \sqrt{1+ \cos(\phi/2)\cos(k)}.
        \end{split}
\end{equation}
In Fig.~\ref{fig:dRGeneral}, the dispersion relation for several values of flux is shown. The flat band $\epsilon_k = 0$ is omitted since it is not modified by the presence of flux.
We introduce here the fermionic operators $f_k^\dagger$, $a_k^\dagger$ and $d_k^\dagger$, which create electrons in the  single-electron
bands, more precisely,  $f_k^\dagger$ creates a particle with momentum $k$ on the top band, $a_k^\dagger$ on the flat band, and $d_k^\dagger$ on the bottom band.
The eigenstates creating operators are given by linear combinations of the creation operators on sites A, B and C, and for the flat band, the expression of the creation operator is rather simple,
\begin{equation}
  \begin{split}
    a_k^\dagger &= \dfrac{1}{ \sqrt{1+\cos \left( \phi/2 \right) \cos(k)} } \left[ \cos\left( \phi/4 - k/2\right) B_k^\dagger \right. \\
		&- \left. \cos \left( \phi/4 + k/2 \right) C_k^\dagger \right].
  \end{split}
  \label{eq:flatstates}
\end{equation}

As can be concluded from the above results, the diamond chain presents always a flux independent dispersionless band and is gapless for zero flux. At finite flux  a gap opens between the bottom and the top bands. This gap is given by
$
    \Delta \epsilon = 4t\sqrt{1-\cos(\phi/2)},
$
while the bandwidth of the system is 
$
    W = 4t\sqrt{1+\cos{(\phi/2)}}.
$

In Fig.~\ref{fig:dRGeneral}, it can be observed that for $k = \pi/2$, two flux independent eigenvalues appear. This condition is only fulfilled if the number of unit cells is a multiple of four, $N_c = 4n$ with $n$ integer.
We should also note that the system is $2\pi$ periodic on the flux when $N_c$ is even and $4 \pi$ periodic when $N_c$ is odd. A flux of $2\pi$ interchanges the energies of the dispersive states with $k=0$ and $k = \pi$. These two values of momenta are allowed when $N_c$ is even. The value $k=\pi$ is forbidden when $N_c$ is even and the period of the system is $4\pi$ on the flux. 

The eigenvalues for an arbitrary value of flux $\phi'$ are obtained following similar steps and 
the energy-dispersion relations become 
\begin{equation}
        \begin{split}
            \epsilon_\text{flat} &= 0, \\
            \epsilon_{\pm} &= \pm 2t \sqrt{1 + \cos(\phi/2)\cos(\phi'/N_c + k)}.
        \end{split}
\end{equation}
The insertion of an extra flux $\phi'$ in the AB$_2$ chain only translates the energy-dispersion relation by $\phi'/N_c$ and therefore does nothing to the flat band. This result is expected since it is similar to what is observed in 1D quantum rings. One can  use $\phi'$ to control the momentum for which the top/bottom band reach its maximum/minimum energy.

\subsection{Density of states, filling and ground-state energy}
Let us consider the thermodynamic limit ($k$ continuous) and calculate the density of states (DOS) of the tight-binding AB$_2$ chain. The DOS of the flat band is  a Dirac-delta function at $\epsilon =0$. 
The dispersive bands are one-dimensional bands and the combined DOS of the two dispersive branches can be written as
\begin{equation}
    D(\epsilon)_\pm = \dfrac{1}{\pi t} \abs{\epsilon'} \dfrac{1}{ \sqrt{ \cos^2(\phi/2)-\left( \epsilon'^2 -1 \right)^2} },
    \label{DOSdispersive}
\end{equation}
when $\epsilon$ belongs to the energy intervals associated with the dispersive bands and where $\epsilon' = \epsilon/2t$.
The full DOS of the system  is  therefore,
$
    D(\epsilon) = D_\pm(\epsilon) + \delta(\epsilon).
$

Filling is defined as the number of electrons $N$ per  site, 
$
    \rho =N/N_s = N/3N_c,
$ 
where $N_s$ is the number of sites.
Due to the symmetric nature of the energy spectrum and since each band can accommodate one electron per unit cell, we know that half filling occurs for $E_F = 0$. Since the flat band contains $N_c$ states then when $\rho \in [1/3,2/3]$ one has $E_F = 0$.
If $E_F$ lies on the bottom band we have
\begin{equation}
    \rho = \dfrac{1}{3} \int_{\epsilon_{b,\text{min}} }^{E_F} D_- (\epsilon) \D \epsilon,
\end{equation}
where $\epsilon_{b,\text{min}} = -2\sqrt{2}t\cos(\phi/4)$ is the bottom of the band and where the factor of 1/3 is due to the fact that the integral of the DOS over a band is one and each band contributes equally to the DOS.
Using Eq. \eqref{DOSdispersive} we have,
\begin{equation}
  \begin{split}
    \rho = \dfrac{1}{6\pi}
        &\left[ 
            \arctan \left( 
                \dfrac{ \epsilon_{b,\text{min}}' }{ \sqrt{\cos^2(\phi/2) - \epsilon_{b,\text{min}}'^2}  } 
            \right) - 
	\right. \\
             & \left.\arctan \left( 
                \dfrac{ E_F' }{ \sqrt{\cos^2(\phi/2) - E_F'^2}  } 
            \right)
        \right],
  \end{split}
        \label{fillingBottom}
\end{equation}
where
\begin{equation}
    \begin{split}
        \epsilon_{b,\text{min}}' &= \left( \dfrac{\epsilon_{b,\text{min}} }{2t} \right)^2 - 1, \\
        E_F ' & = \left( \dfrac{E_F}{2t} \right)^2 - 1.
    \end{split}
    \label{fillingTop}
\end{equation}
If, on the other hand, $E_F$ lies on the top band, one can use the symmetry of the density of states to write $\rho=1- \rho_h$, where 
\begin{equation}
    \rho_h = \dfrac{1}{3} \int_{\epsilon_{b,\text{min}} }^{- E_F } D_- (\epsilon) \D \epsilon,
\end{equation}
is given by Eq.~\eqref{fillingBottom}.

\begin{figure}[t]
    \centering
    \includegraphics[height=5cm]{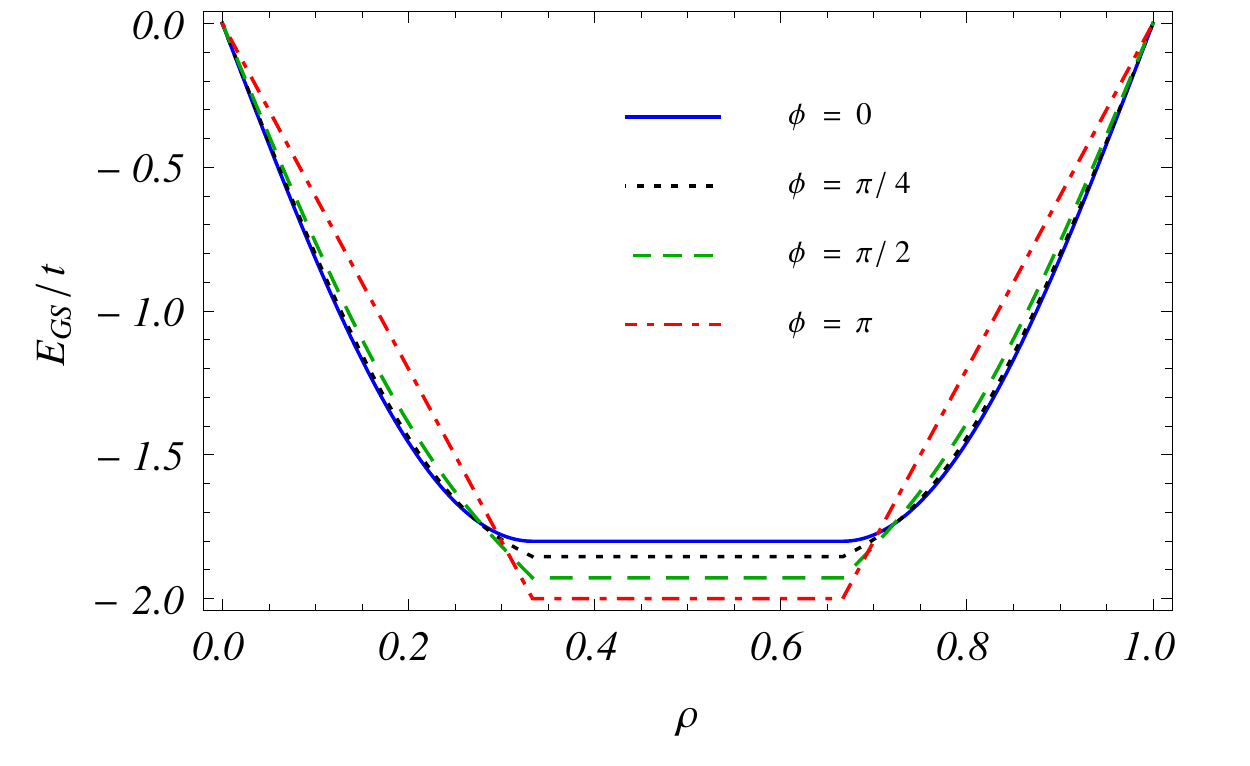}
    \caption{Ground state energy in the thermodynamic limit as a function of filling for several values of flux. $E_{\text{GS}}$ remains constant when  filling the flat band, $\rho \in [1/3,2/3]$ achieving its minimum value in that interval for $\phi = \pi$. For $\phi = \pi$, $E_{\text{GS}}$ has linear behavior in $\rho$, since all bands are flat.}
    \label{fig:energyGS}
\end{figure}
Given the preceding results, we are able to calculate the ground-state energy of the system as a function of  filling. The ground-state energy is given by
\begin{equation}
    E_{\text{GS}} = \int^{E_F}_{E_{\text{min}}} D(\epsilon) \epsilon \D \epsilon,
\end{equation}
while Eqs. \eqref{fillingBottom} and \eqref{fillingTop} give us the relation between the Fermi energy and the filling. For a general value of $\phi$ there is no simple analytical expression and the ground-state energy should be calculated numerically. However, when $\phi = \pi$ it is very simple to derive an exact result, given that the system only has three energy levels, $\epsilon = 0, \pm 2t$ and that each can take $N_c$ electrons. In this case, the ground-state energy per unit cell while the bottom band is not fully filled is given by $E_{\text{GS}} = -2t N/N_c = -6t \rho$. At $\rho = 1/3$ the bottom band gets fully filled and from $\rho = 1/3$ to $\rho = 2/3$ we are filling the middle band whose energy is $\epsilon = 0$. Therefore, $E_{\text{GS}} = -2t, \quad \rho \in [1/3, 2/3]$. For $\rho > 2/3$ we start filling the upper band, whose energy is $\epsilon = 2t$. In this situation the ground-state energy is given by $E_{\text{GS}} = 6t(\rho - 2/3) - 2t$. 

In Fig. \ref{fig:energyGS} we plot the ground-state energy as a function of filling for several values of flux. As can be seen for $\rho \in [1/3,2/3]$ the ground-state energy remains constant since we are filling the flat band $\epsilon_k = 0$. The ground-state energy is obviously even around $\epsilon_k = 0.5$ due to the symmetric nature of the dispersion relation. We also see that for $\rho \not\in [1/3,2/3]$ the dependence of the ground-state energy on the filling departs from its linear behavior when $\phi \neq \pi$.

\section{Flat bands and localized states for arrays of quantum rings}

It is well known that the single-particle flat band eigenstates of a  geometrically frustrated lattice can be written as a set of localized eigenstates which are  translated versions of the same state $\vert \psi_{\text{loc}} \rangle$ \cite{}.
This single-particle state is non-zero in a small lattice region and it is an eigenstate of the tight-binding Hamiltonian (a zero energy eigenstate in the case of the AB$_2$ chain). This means that if we write the eigenstate as $\vert \psi_{\text{loc}} \rangle =\sum_i a_i c_i^\dagger \vert 0 \rangle$, where $i$ runs over all lattice sites such that $a_i \neq 0$, then $\sum_i t_{ri} a_i=\epsilon a_r=0$, where $r$ can be any site in the lattice such that  $a_r = 0$, $t_{ri}$ is the hopping constant between sites $r$ and $i$ and $\epsilon$ is the kinetic energy of the state. In the particular case of AB$_2$ chains, $\epsilon=0$ and the restrictions on sites $i$ and $r$ can be lifted, i.e., 
$\sum_i t_{ri} a_i=\epsilon a_r=0$, where $i$ runs over all lattice sites and $r$ is any site of the lattice.
We propose a particular perspective (but equivalent) for the case of any array of quantum rings similar to the AB$_2$ chain which easily allows us to identify the localized states in the absence or presence of flux.
Let us also assume for now that the array is such that the shared sites between consecutive rings are directly opposite, and therefore the number of sites in each ring is even.

\begin{figure}[t]
    \includegraphics[width=8cm]{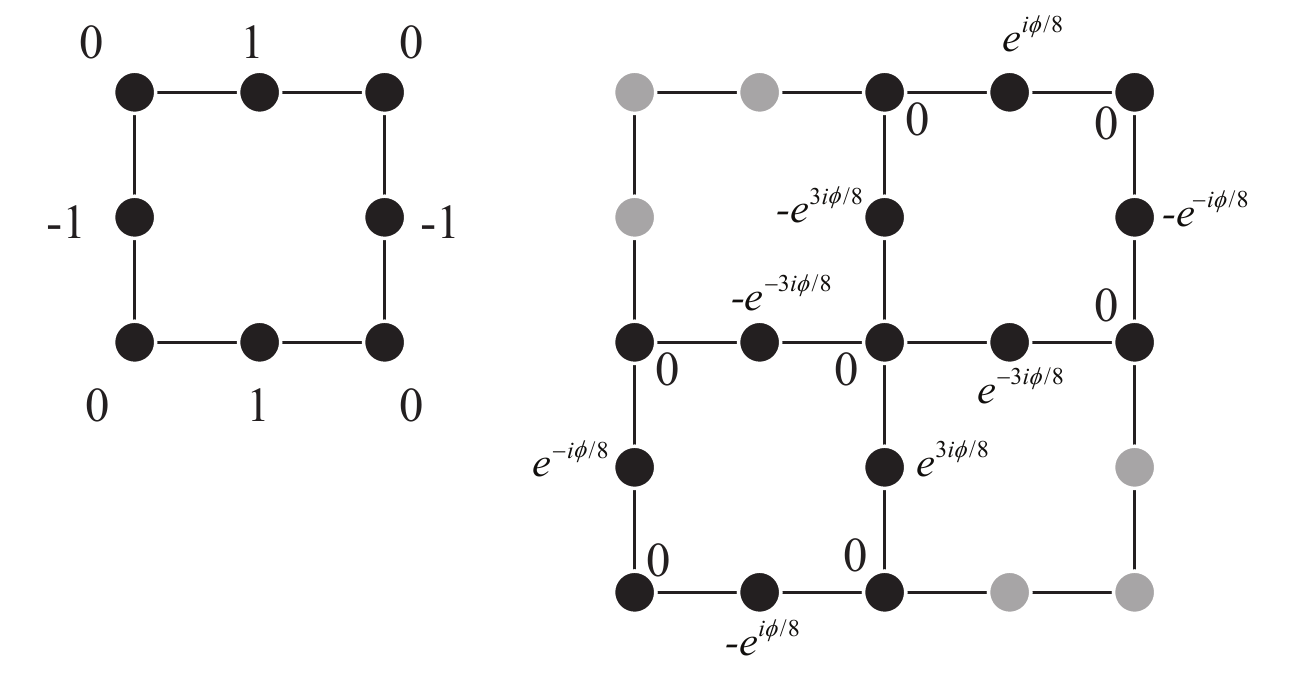}
    \caption{Localized states in the case of the Lieb lattice with or without magnetic flux. The introduction of magnetic flux extends the localized state to two plaquettes. These plaquettes are equivalent to the quantum rings discussed in the text and the localized state is a standing wave with nodes in the plaquettes vertices.}
    \label{fig:squarefrustrated1}
\end{figure}

For zero flux, $\phi=\phi_o=\phi_i=0$,  all energy levels of the tight-binding ring are doubly degenerate with exception of the lowest and highest energy levels ($k=0$ and $k=\pi$) and the respective eigenstates have opposite momenta, reflecting the time-reversal symmetry of the Hamiltonian.
In the particular case of the AB$_2$ chain, the respective ring has four sites and there is only one degenerate level corresponding to zero energy. Subtracting the states of opposite momenta, one obtains a standing wave with nodes at $i=0$ and $i=N_r/2$ (ring sites are numbered clockwise from $i=0$ to $i=N_r-1$).
For $N_r > 2n$ with $n>2$ one has more than one flat band. More precisely, one has $N_{\text{ring}}/2-1$ flat bands with energies $\epsilon_n=-2t \cos (2 \pi n/N_r)$ with $n=1, \ldots, N_r/2-1$.
Since these nodes coincide with the sites shared by consecutive rings, one can say that for these standing wave states, the ring becomes uncoupled to the rest of the chain. Furthermore, if $N_r > 2n$  with $n>2$ the standing waves may have more than two nodes and more complex lattices can have such localized states. Note however that the nodes position must be an integer. We conclude  that in order to find  flat band eigenstates of arrays of quantum rings, one may construct standing waves such that the nodes coincide with sites shared between different quantum rings.
For instance, for $N_r=8$, a square lattice is possible as shown in Fig.~\ref{fig:squarefrustrated1}.

These arguments agree with the results obtained for the AB$_2$ chain.  For $\phi=\phi_o=\phi_i=0$, one has 
$
    a_k^\dagger =  [ B_k^\dagger -   C_k^\dagger ]/\sqrt{2},
$
which can be Fourier transformed to obtain
$
    a_i^\dagger = [ B_i^\dagger -   C_i^\dagger ]/\sqrt{2}.
$
The single-particle state $ a_i^\dagger \vert 0 \rangle$ is a localized eigenstate in cell $i$ with zero energy.

Let us consider now $\phi=0$ but $\phi_o=-\phi_i\neq 0$. Then the tight-binding constants in one ring become $t_{j,j+1}=e^{-i\phi_o/2N_c \cdot j} t$, for $j=0, \ldots,N_r/2-1$ and 
$t_{j,j+1}=e^{i\phi_o/2N_c \cdot j} t$, for $j=N_r/2, \ldots,N_r-1$ so that the total Peierls phase is zero reflecting the fact that $\phi=0$. A simple gauge transformation
\begin{equation}
\begin{array}{ccll}
	c_j^\dagger & \rightarrow & e^{i\frac{\phi_o}{2N_c} \cdot j} c_j^\dagger, 
	&\quad j=0, \ldots,\frac{N_r}{2} \\
	c_j^\dagger & \rightarrow & e^{i \frac{\phi_o}{2N_c} \cdot (N_r- j)} c_j^\dagger, 
	&\quad j=\frac{N_r}{2}+1, \ldots,N_r-1 \nonumber
\end{array}
\end{equation}
restores the translation invariance with a zero Peierls  phase and the previous construction of localized states can be applied. Therefore, the same flat bands will be present.  In the case of the AB$_2$ chain, the flat-band states will have the form 
$
    a_i^\dagger \vert 0 \rangle = e^{-i \frac{\phi_o}{2N_c}} [ B_i^\dagger -   C_i^\dagger ]/\sqrt{2}  \vert 0 \rangle,
$
but the phase term is obviously irrelevant.

\begin{figure}[t]
    \includegraphics[width=8cm]{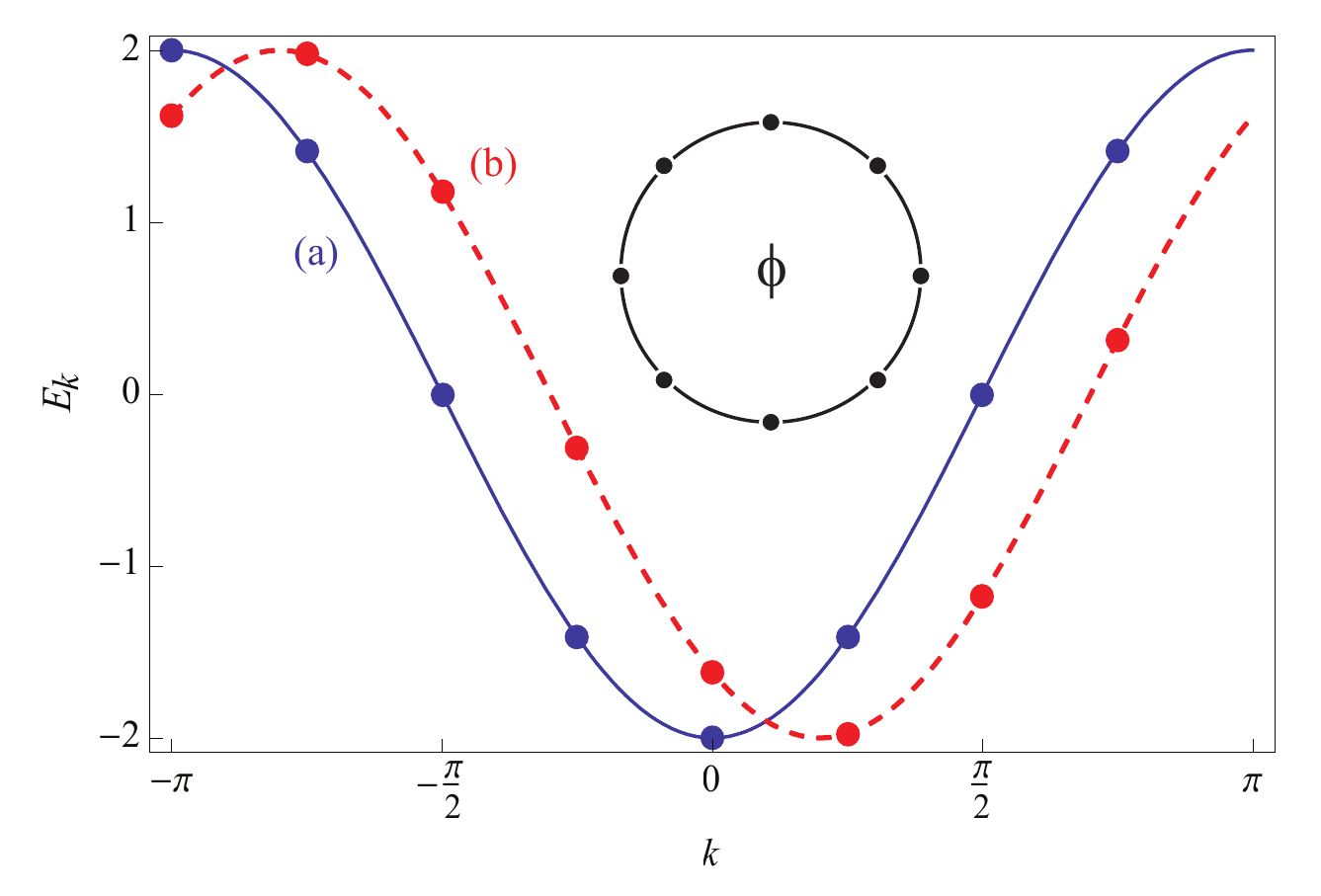}
    \caption{Dispersion relation of a tight-binding ring with eight sites: (a) without flux; (b) threaded by a magnetic flux  $\phi=\pi/5$. For zero flux,  states $k$ and $-k$ are doubly degenerate (with the obvious exception of $k=0$ and $k=\pi$). When $\phi\neq 0$, the eigenstates of the tight-binding ring with opposite momenta $k$ and $-k$ do not have the same energy reflecting the loss of time-reversal symmetry.}
    \label{fig:ring}
\end{figure}
\begin{figure}[t]
    \includegraphics[width=8cm]{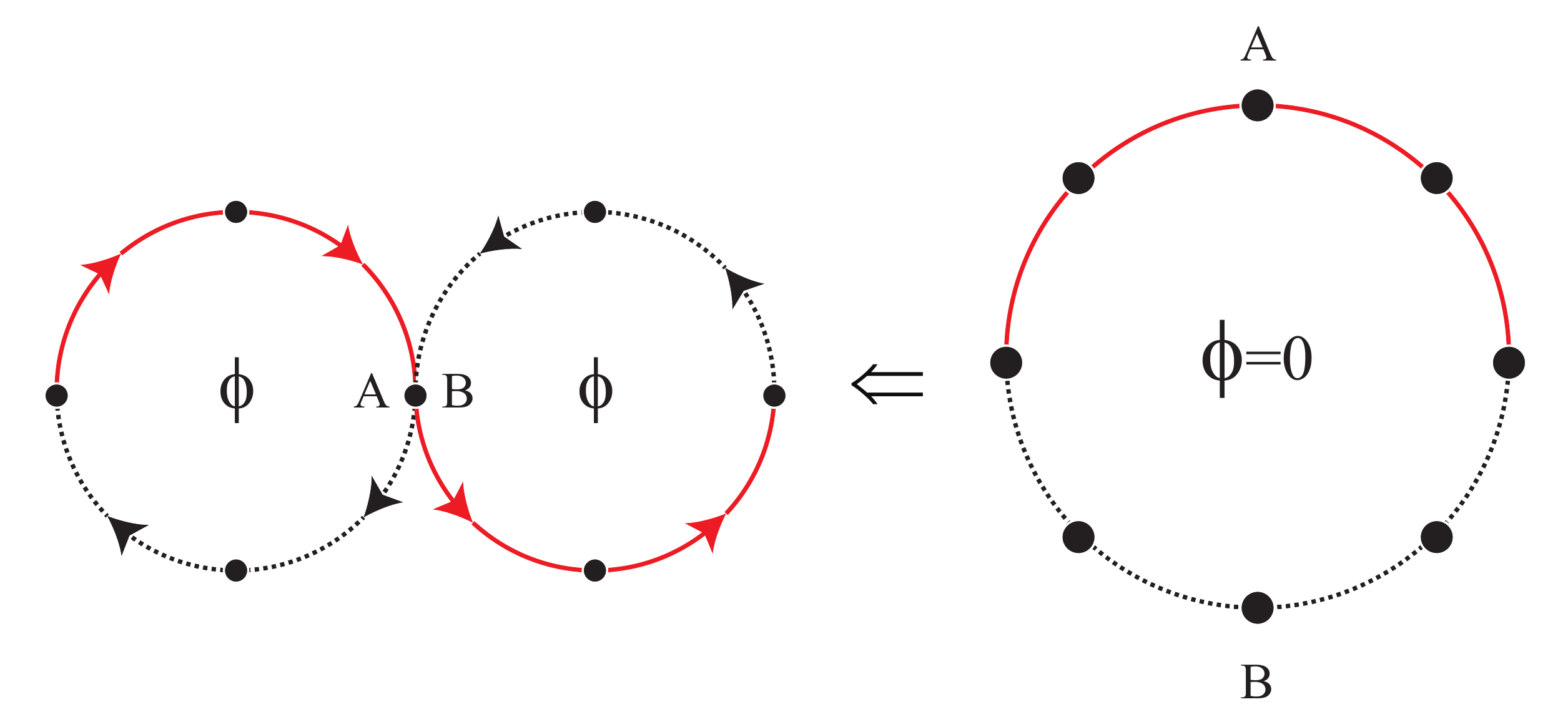}
    \caption{In the presence of flux, considering two consecutive rings and following  first the continuous path (in red) and then the dotted path (in black) the total Peierls phase in that path is zero (the path is clockwise for the left ring and counter-clockwise for the right ring). If one constructs standing waves for a ring of $2 N_r$ sites with the additional condition of extra nodes at sites $A$ and $B$, this state can be folded to give an eigenstate of the two rings threaded by flux with nodes at the extremities and at site A. }
    \label{fig:tworings}
\end{figure}

Let us address now the case of $\phi\neq 0$ but $\phi'=0$ (the case with $\phi'\neq 0$ is obtained following the same steps after a gauge transformation as explained for the $\phi=0$ case). When $\phi\neq 0$, eigenstates of the tight-binding ring with opposite momenta $k$ and $-k$ do not have the same energy (see Fig.~\ref{fig:ring}) and therefore the standing wave obtained from the subtraction of these states is not an eigenstate of the tight-binding Hamiltonian of the ring. One can overcome this difficulty considering two consecutive rings and following a certain path in those rings such that the total Peierls phase in that path is zero as shown in Fig.~\ref{fig:tworings}. If one constructs standing waves for a ring of $2 N_r$ sites with the additional condition of extra nodes at sites $N_r/2$ and $2 N_r- N_r/2$ (sites $A$ and $B$ in Fig.~\ref{fig:tworings}), this state can be folded (making site $A$ overlap with site $B$) to give an eigenstate of the two rings threaded by flux with nodes at the extremities. Note that the larger ring has non-zero Peierls phases associated with each hopping but the total phase is zero. So after a gauge transformation to eliminate these phases, the states of opposite momenta are degenerate and the standing wave is an eigenstate of the Hamiltonian of the two rings.
Again, one or more flat bands are possible depending on the size of the quantum ring.

The gauge transformation for the larger ring will be
\begin{equation}
\begin{array}{ccll}
	c_j^\dagger & \rightarrow & e^{i\frac{\phi_o}{2N_c} \cdot j} c_j^\dagger, 
	&\, j=0, \ldots,\frac{N_r}{2} \\
	c_j^\dagger & \rightarrow & e^{i \frac{\phi_o}{2N_c} \cdot (N_r- j)} c_j^\dagger, 
	&\, j=\frac{N_r}{2}+1, \ldots,N_r \nonumber \\
	c_j^\dagger & \rightarrow & e^{-i\frac{\phi_o}{2N_c} \cdot (j-N_r)} c_j^\dagger, 
	&\, j=N_r+1, \ldots,N_r+\frac{N_r}{2} \nonumber \\
	c_j^\dagger & \rightarrow & e^{-i \frac{\phi_o}{2N_c} \cdot (2N_r- j)} c_j^\dagger, 
	&\, j=N_r+\frac{N_r}{2}+1, \ldots,2N_r-1  \nonumber 
\end{array}
\end{equation}
where the sites have been numbered clockwise in the larger ring.
In the case of the AB$_2$ chain, after the gauge transformation the localized state (not normalized) will be 
$
	(c_1^\dagger-c_3^\dagger+c_5^\dagger-c_7^\dagger) \vert 0 \rangle,
$
which, inverting the gauge, transformation corresponds to the state 
$
	[e^{-i\frac{\phi_o}{2N_c}} (c_1^\dagger- c_3^\dagger)
	+ e^{i\frac{\phi_o}{2N_c}}( c_5^\dagger- c_7^\dagger)] \vert 0 \rangle
$.
This state  written in terms of the operators of the AB$_2$ chain becomes
$
	[e^{-i\frac{\phi_o}{2N_c}} (B_j^\dagger- C_{j+1}^\dagger)
	+ e^{i\frac{\phi_o}{2N_c}}( B_{j+1}^\dagger- C_j^\dagger)] \vert 0 \rangle
$
or equivalently 
$
	[e^{-i\frac{\phi}{4}} (B_j^\dagger- C_{j+1}^\dagger)
	+ e^{i\frac{\phi}{4}}( B_{j+1}^\dagger- C_j^\dagger)] \vert 0 \rangle
$.
These are exactly the localized eigenstates  obtained Fourier transforming Eq.~\eqref{eq:flatstates},
\begin{equation}
  \begin{split}
    2 e^{i k/2}  & \sqrt{1+\cos \left( \phi/2 \right) \cos(k)} a_k^\dagger =  \\
    & [e^{i\phi/4}+ e^{-i(\phi/4 - k)}] B_k^\dagger 
	- [e^{i(\phi/4 + k)}+ e^{-i\phi/4}]  C_k^\dagger .
  \end{split}
\end{equation}
 Note that these localized states may overlap in real space, that is, they constitute a basis of the subspace of localized states but not an orthogonal basis. Such impossibility of constructing orthogonalized Wannier states for certain lattices with flat bands under a uniform magnetic field has been mentioned before \cite{Aoki1996}.

As an example of application of the previous arguments, we show in Fig.~\ref{fig:squarefrustrated1} the localized states for the Lieb lattice. It is known that this lattice displays a flat band for zero and finite magnetic flux. Aoki {\it et al}. have found the respective localized state for zero flux as well as by inspection a localized  ``elongated ring state'' for finite flux \cite{Aoki1996}. The zero flux localized state agrees with the one shown in Fig.~\ref{fig:squarefrustrated1}, but our localized state for finite flux shown in Fig.~\ref{fig:squarefrustrated1} is considerably more compact than the ``elongated ring state'' of Ref.~\cite{Aoki1996}.
We expect that the previous arguments can be applied to other lattices that fall onto the Lieb lattice category,  that is,  bipartite lattices with  different numbers, $n_A$ and $n_B$, of A and
B sublattice sites in a unit cell \cite{Lieb1989}. 
Another example of simple application is the $B_c$ class superhoneycomb lattice \cite{Aoki1996}.  Again our argument justifies the fact that the flat band in this system remains flat for finite magnetic field.

\section{Mean-field results for general V}
In this section, we present a mean-field study of the t-V AB$_2$ chain taking into account nearest-neighbor Coulomb interaction. The results obtained must be interpreted with  caution, having in mind the known drawbacks of this approach. One of these drawbacks is the fact that this approach neglects thermal fluctuations, which as stated by the Mermin-Wagner theorem prevent long range order at finite temperature. In this section, only the zero-temperature case is addressed, so one avoids this problem.
Even at zero temperature, the mean-field approach overestimates the existence of an ordered phase, since quantum fluctuations oppose the emergence of an ordered phase and these quantum fluctuations are particularly strong in quasi-1D systems. Note however that the ground state of quasi-1D systems may in fact be ordered despite these quantum fluctuations.

When considering nearest-neighbor Coulomb interactions, we will be interested on the density of particles at A, B and C sites. We assume the particle density on site X to be the average number of particles per number of unit cell, $\rho_X = \meanline{N}_X/N_c$. We also assume that the particle density at B and C sites is the same. Let $\rho$ denote the total particle density,  
\begin{equation}
    \rho = \dfrac{\rho_A + 2 \rho_B}{3},
\end{equation}
where $\rho_A$ and $\rho_B$ are the particle densities at A and B sites respectively. We then have
\begin{equation}
    0 \leq
    \begin{Bmatrix}
        \rho_{\phantom{A}} \\
        \rho_A \\
        \rho_B
    \end{Bmatrix}
    \leq 1.
\end{equation}
In this situation, the interaction part of the mean-field Hamiltonian can be written as
\begin{equation}
    H_{\text{int}} = 2V \sum_j \left[ 2 \rho_B n_j^A + \rho_A n_j^B + \rho_A n_j^C \right] - 4N_c\rho_A\rho_B.
\end{equation}
While there is no simple expression for the mean-field dispersion relation for general $\phi$, a simple expression exists for zero flux,
\begin{equation}
    \begin{split}
        \epsilon_\text{flat} &= 2 V \rho_A ,\\
        \epsilon_{\pm} &= V \left( \rho_A + 2\rho_B \right) \pm \sqrt{8t^2\cos^2(k/2) + \Delta_V^2},
        \label{eq:meanfield_eq}
    \end{split}
\end{equation}
where $\Delta_V = V \left(\rho_A - 2\rho_B\right)$. Again, a flat band is present, but its energy level depends on the density at sites A. 

We define the order parameter as the excess of density at the A sites,
\begin{equation}
    \Delta \rho = \dfrac{\rho_A - 2\rho_B}{\rho_A + 2\rho_B}.
\end{equation}
Due to the equivalence of the B and C sites, we regard our system as being a dimerized system consisting of two alternating types of sites: A sites and BC pseudo-sites. Note however that this picture is to be interpreted with caution since BC pseudo-sites are not real sites and can accommodate twice as many electrons as  A sites.
One must note that in a general situation we have $\Delta \rho \in [-1,1]$. Since $ 0 < \rho_A,\rho_B < 1$, then for $\rho > 1/3$, $\Delta \rho$ is limited to the interval $[-1,\Delta \rho_{\text{max}}]$, where $\Delta\rho_{\text{max}}$ lies in $]-1,0[$ and whose value decreases with increasing $\rho$. For the same reason, for $\rho > 2/3$, $\Delta \rho$ is limited to $]-1,-1/3]$, and not only the upper limit of $\Delta \rho$ decreases with increasing $\rho$, but also the lower limit of $\Delta \rho$ increases with increasing $\rho$. For $\rho = 1$ we have $\Delta \rho = -1/3$, corresponding to an equal density of particles on every site (the only possible state when the system is completely filled).
\begin{figure}[t]
    \centering
    \includegraphics[width=8cm]{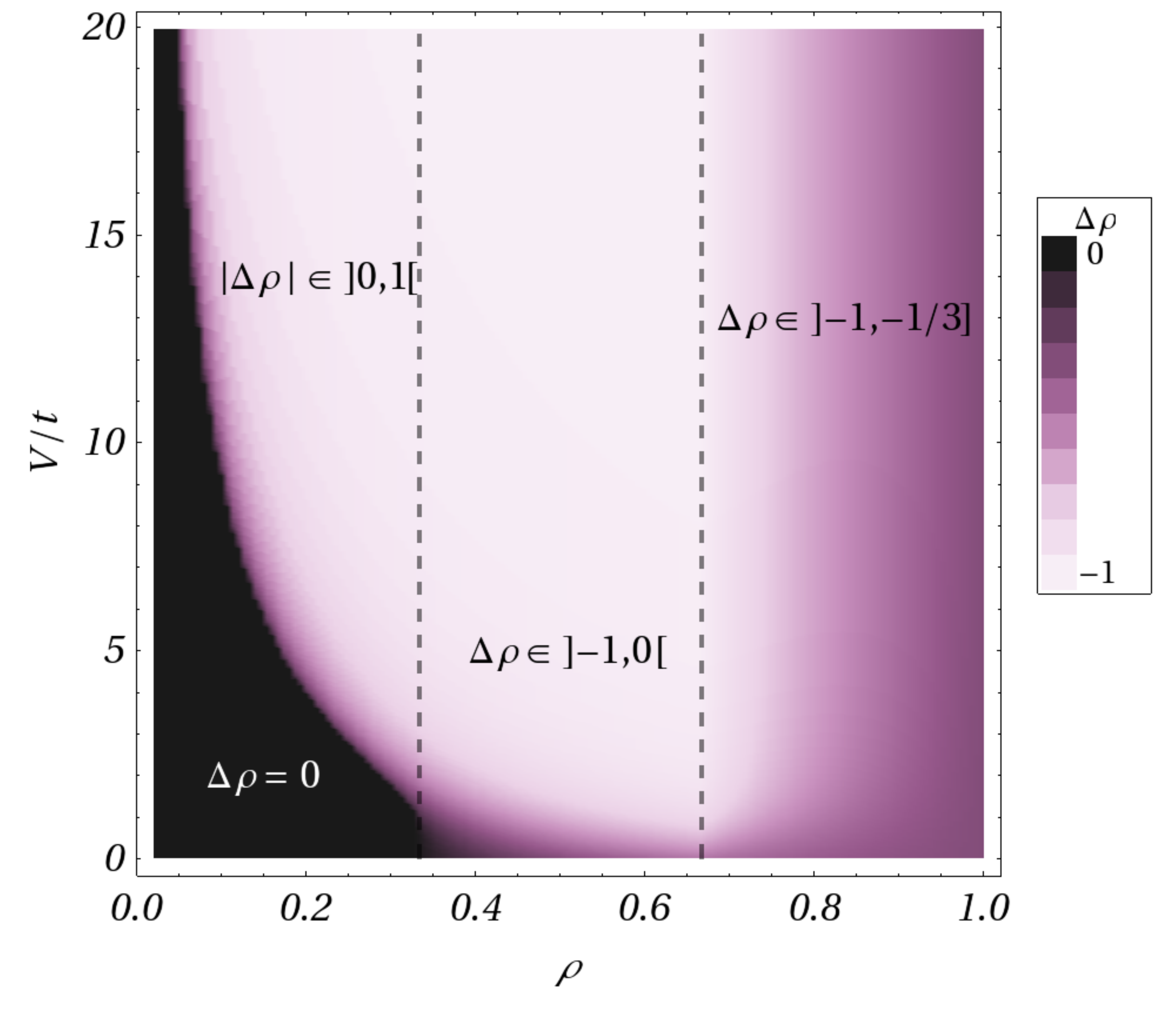}
    \caption{$V/t$ versus filling mean-field phase diagram for the spinless diamond chain considering  nearest-neighbor Coulomb interactions and $N_c = 128$. For $\rho < 1/3$ we can have a uniform density phase ($\Delta \rho =0$) or a phase with excess of density at A or BC pseudo-sites ($\Delta \rho >0$ or $\Delta \rho < 0$, respectively). For $\rho > 1/3$, Pauli's exclusion principle breaks the symmetry between A and BC pseudo-sites. A uniform density phase is not possible anymore and the density of particles can no longer be situated only at A sites. It can however, for $\rho < 2/3$, be situated only at BC pseudo-sites. For $\rho > 1/3$, due to Pauli's exclusion principle, the density of particles is required to be spread among A and BC sites and for $\rho = 1$ the order parameter is $\Delta \rho = -1/3$, which implies a uniform density of particles between the real A, B and C sites.}
    \label{fig:Vphase}
\end{figure}
\begin{figure}[t]
    \centering
    \includegraphics[width=8cm]{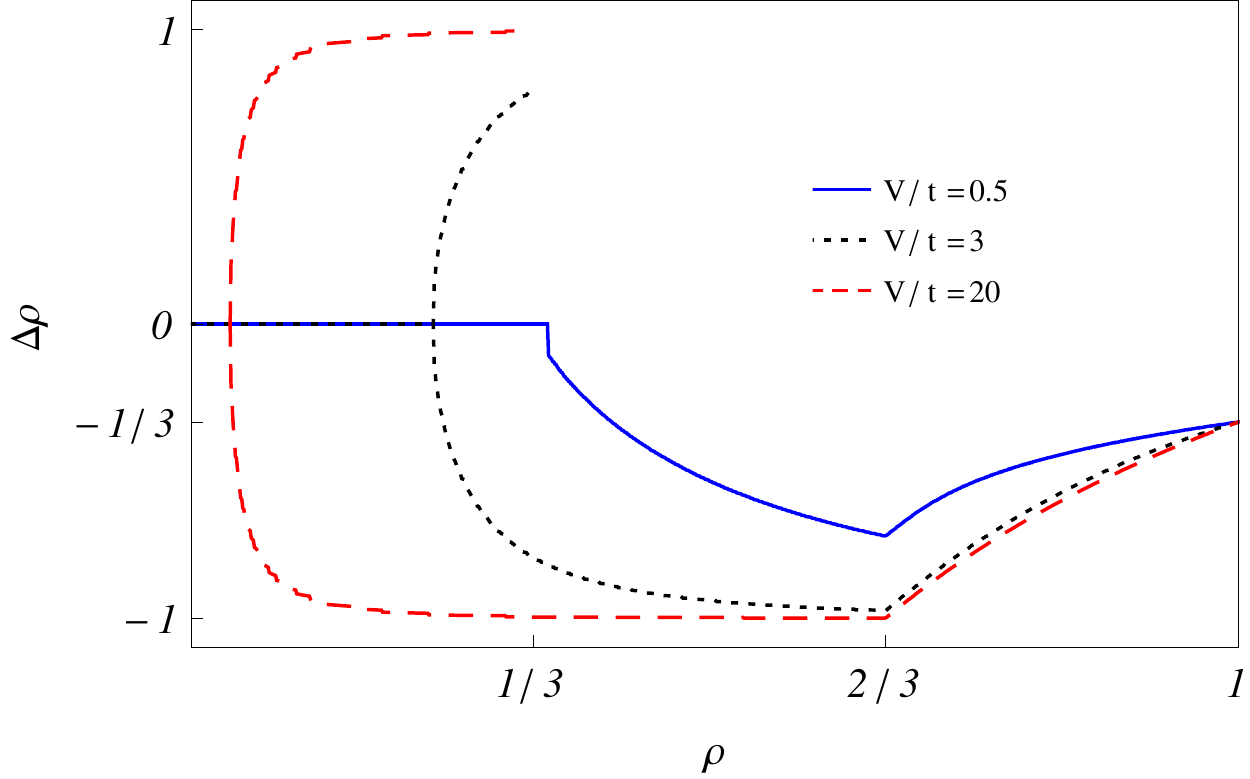}
    \caption{Order parameter as a function of filling for $V/t=0.5$, 3 and 20 and $N_c = 200$. The changes of slope of the order parameter indicate the phase transitions. Note that the absence of a positive solution of $\Delta \rho$ for $\rho > 1/3$ reflects the fact that the A and BC pseudo-sites may be occupied by one and two fermions, respectively. As a consequence, for $\rho < 1/3$ the order parameter can be double valued since we are able to localize all fermions at A sites ($\Delta \rho > 0$) or at BC pseudo-sites ($\Delta \rho < 0$). Since for $\rho > 1/3$ we can no longer localize all fermions at A sites, the order parameter can no longer be positive.}
    \label{fig:deltan}
\end{figure}

The phase diagram of the system is depicted in Fig. \ref{fig:Vphase}. a uniform density phase  can exist only for a filling $\rho < 1/3$. For $\rho < 1/3$, starting in the uniform density phase, by increasing the interaction we are able to localize an excess of electron density at A sites or at BC pseudo-sites, both situations being symmetric. By further increasing the interaction we are able to localize all the fermions at A sites or at BC pseudo-sites where both situations remain symmetric. As a consequence the order parameter $\Delta \rho$ can be double valued for these values of filling. In the region $1/3 < \rho < 2/3$ one no longer can localize the full electrons density at A sites while one can at BC sites and therefore although the Hamiltonian treats A and BC pseudo-sites equally, Pauli's exclusion breaks the symmetry between A and BC pseudo-sites, lowering the symmetry of the system. Consequently, for $\rho > 1/3$, the order parameter can no longer be double valued. In the region $\rho > 2/3$, again due to Pauli's exclusion principle, one is not able to fully localize the density of electrons even at BC pseudo-sites and only one phase remains.

In Fig.~\ref{fig:deltan}, the order parameter as a function of filling  is shown for $V/t=0.5$, 3 and 20 and $N_c = 200$. The changes of slope of the order parameter indicate the phase transitions. The absence of a positive solution of $\Delta \rho$ for  $\rho > 1/3$ reflects the fact that the  A and BC pseudo-sites may be occupied by one and two fermions, respectively.
\begin{figure}[t]
    \centering
    \includegraphics[width=8cm]{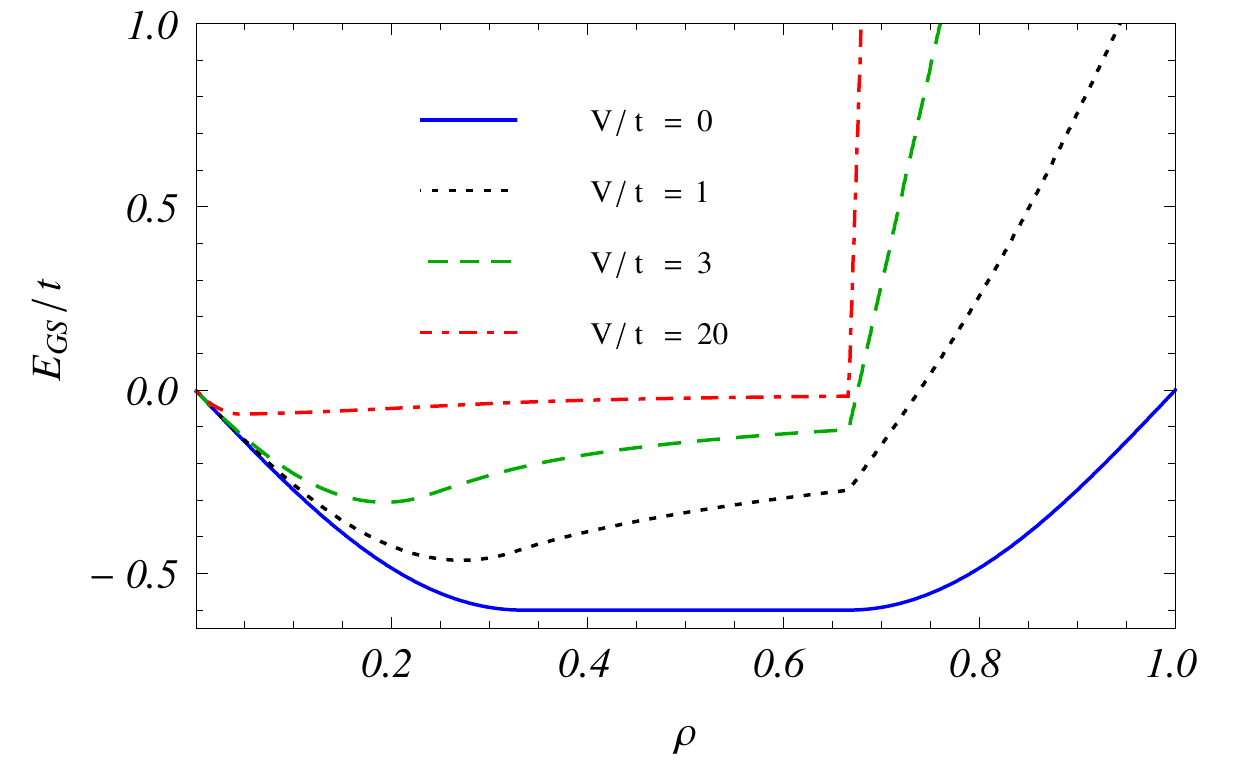}
    \caption{Ground state energy per site as a function of $\rho$ for $V/t=0$, 1, 3 and 20 and $N_c = 200$, obtained in the mean-field approach. For finite $V$, no flat region appears and the minimum energy is shifted to lower filling. The ground-state energy remains negative for $\rho<2/3$. For $\rho>2/3$, the ground-state energy is almost linear in the filling with a large slope, since nearest-neighbor pairs are being created.  }
    \label{fig:EGS_v}
\end{figure}

In Fig.~\ref{fig:EGS_v}, we show the ground-state energy per site as a function of $\rho$ for $V/t=0$, 1, 3 and 20 and $N_c = 200$, obtained in the mean-field approach. For finite $V$, no flat region appears and the minimum energy is shifted to lower filling.  Recalling Eq.~\eqref{eq:meanfield_eq}, one concludes that a constant  term $3V \rho^2$ is added to the non-interacting ground-state energy when $\Delta \rho=0$, which shifts  the minimum energy  to lower  $\rho$ when  $V$ is small ($\Delta \rho=0$ for low  $\rho$).
The ground-state energy remains negative for $\rho<2/3$. For $\rho>2/3$ and large $V$, the ground-state energy is almost linear in the filling with a large slope, since nearest-neighbor pairs are being created. 

\section{The AB$_2$ chain in the strong-coupling limit $V \rightarrow \infty$}
Making $t/V$ a small parameter, one drives the t-V AB$_2$ model into the so-called strong-coupling limit. There are two equivalent ways to reach this limit, either increasing $V$ or reducing $t$. If $t=0$, the fermions are localized  and all states 
with the same number of 
pairs of nearest-neighbor occupied sites, $\sum_i (n^A_j + n^A_{j+1})( n^B_j + n^C_j)$, 
are degenerate. 
This degeneracy is much lower compared with the ground-state degeneracy of the $U/t \rightarrow \infty$ Hubbard AB$_2$ model.
This degeneracy is lifted if $t/V$ is finite and
up to first order in $t$, the eigenvalues are obtained diagonalizing the 
Hamiltonian within each of the degenerate subspaces. The Hamiltonian within each subspace is obtained using the Gutzwiller projection operators $P_l$, which project onto the subspace with $l$ pairs of nearest-neighbor occupied sites.
The set of eigenstates and
eigenvalues of $l=0$ subspace of this model can be determined 
as we will show below, relying at least for some of the eigenstates in the knowledge of the solution of the t-V chain \cite{Dias2000}. 

In order to simplify the comparison between the t-V chain and the AB$_2$ chain, we number the sites in a different way, using odd numbers for A sites and even numbers for B and C sites.
\begin{widetext}
Let us also define the operator $n^h_i=(1-c_{i}^\dagger c_{i})$ where $c$ can be an operator $A$, $B$ or $C$.
In the strong-coupling limit $ V/t \gg 1$, we obtain  for the ground state subspace ($l=0$)
the projected Hamiltonian with $\phi=0$ (but $\phi_i=\phi_o \neq0$)
\begin{equation}
    P_0 H P_0 = 
    -t \sum_{j\; \text{odd}}  
    \e{\I \phi_o/2N_c}  
    \left[ 
    \prod_{ i \; \epsilon \;\mathcal{P}_j} 
    n^{h}_{i}   \times 
    A_j^\dagger (B_{j+1} +  C_{j+1})   +  
    \prod_{ i \; \epsilon \;\mathcal{P}_{j+1}} 
    n^{h}_{i}   \times 
   (B_{j+1}^\dagger +  C_{j+1}^\dagger) A_{j+2}
    \right] 
    + \text{H.c.} \nonumber
\end{equation}
where $\mathcal{P}_j$ is the set of sites nearest-neighbors of the pair of sites $j$ and $j+1$ (excluding these sites)
and the product of hole occupation numbers reflects the condition that a nearest-neighbor pair of occupied sites is not created when a particle hops.
\end{widetext}
Again, we emphasize that states belonging to subspaces with $l\neq 0$ pairs of nearest-neighbor occupied sites will be discarded since their energy is of the order of $V$. 

\begin{figure}[t]
    \centering
    \includegraphics[width=5cm]{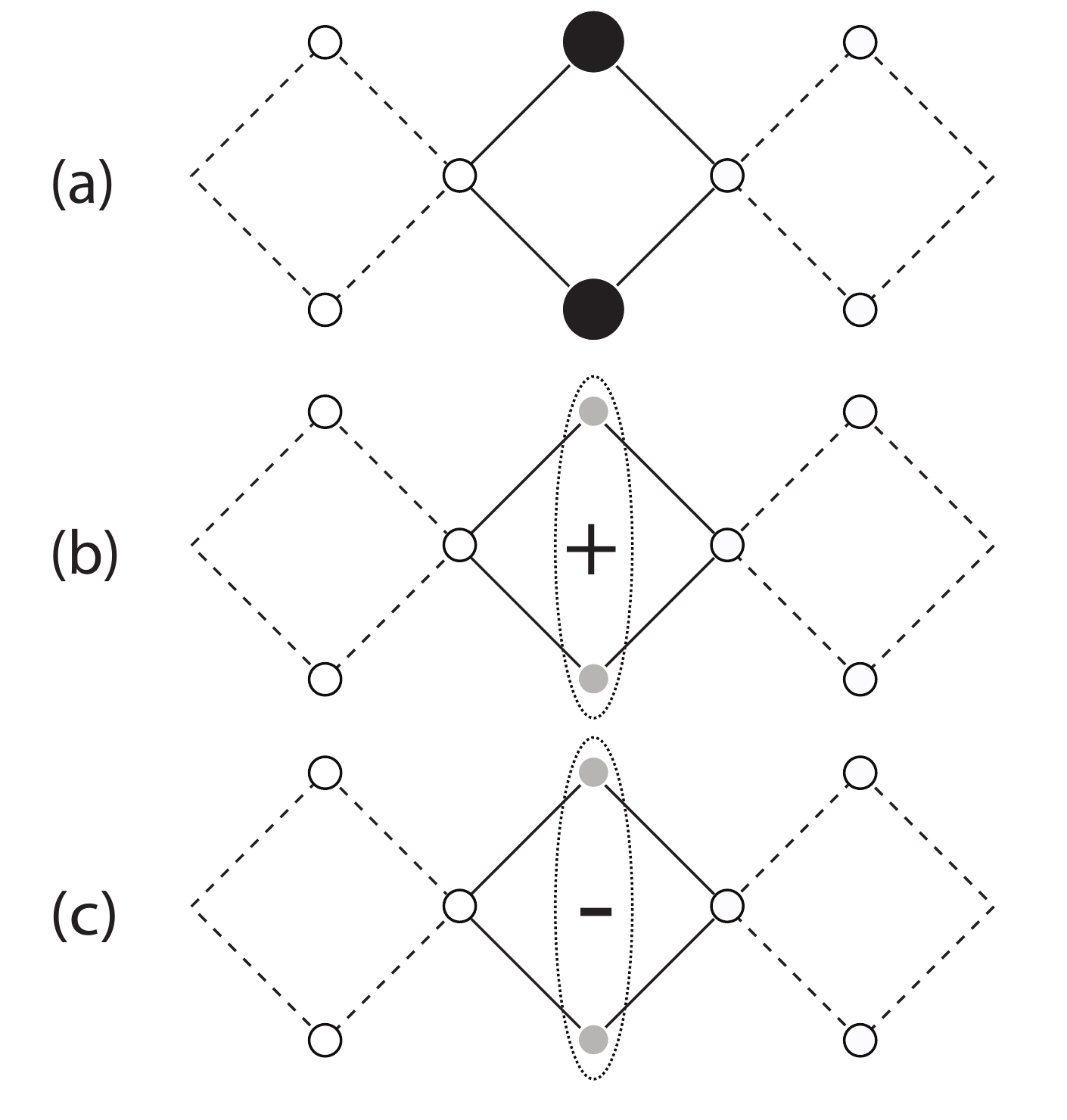}
    \caption{(a) Two-particle localized state and one-particle  (b) itinerant and (c) localized  state involving B and C sites. The filled and empty circles indicate occupied and unoccupied sites respectively. Gray circles indicate bonding (+) and anti-bonding (-) occupations.  The particles in (a) are localized due to the interaction ($V/t =\infty$) and in (c) due to geometric frustration. In (b), the particle is free to hop to the neighboring $A$ sites.}
    \label{fig:statesBC}
\end{figure}
Let us consider two consecutive sites and therefore, nearest neighbors of
each other. Considering the subspace with $l = 0$,
there are three possible configurations for this pair of sites,
which we will call links and they are
$
         (h \, p); (p  \, h); (h  \, h)
$
where $p$ stands for an occupied site and $h$ for an empty one.
The total number of these links in the AB$_2$ chain is given by
\begin{equation}
        N_{hp}+N_{ph}+N_{hh}=4 N_c 
\end{equation}
and $N_{hp}=N_{ph}$. Note that unlike the case of the t-V chain, the number of
links  $N_{hp}$
and $ N_{hh}$ in the strong coupling AB$_2$ chain is not a conserved quantity.

\subsection{Basis}

Let us consider a unit cell of the AB$_2$ chain (which has three sites, A, B and C).
The set of states for this cell correspond to five possible configurations in what concerns the particles distribution :
i) zero particles with all sites being unoccupied; ii) one particle which may be at site A,B or C. The states with just one particle either at site B or site C can be combined to give a bonding and an anti-bonding state [see Fig.~\ref{fig:statesBC}(b) and Fig.~\ref{fig:statesBC}(c)]. The anti-bonding state is a localized state as discussed in previous sections. The bonding state is a itinerant state; iii) two particles which must be at sites B and C in order to avoid a nearest neighbor occupied pair of sites [see Fig.~\ref{fig:statesBC}(a)]. These particles are also localized particles because if they hopped, a state with a nearest neighbor occupied pair of sites would be created.

\begin{figure}[t]
    \centering
    \includegraphics[width=9cm]{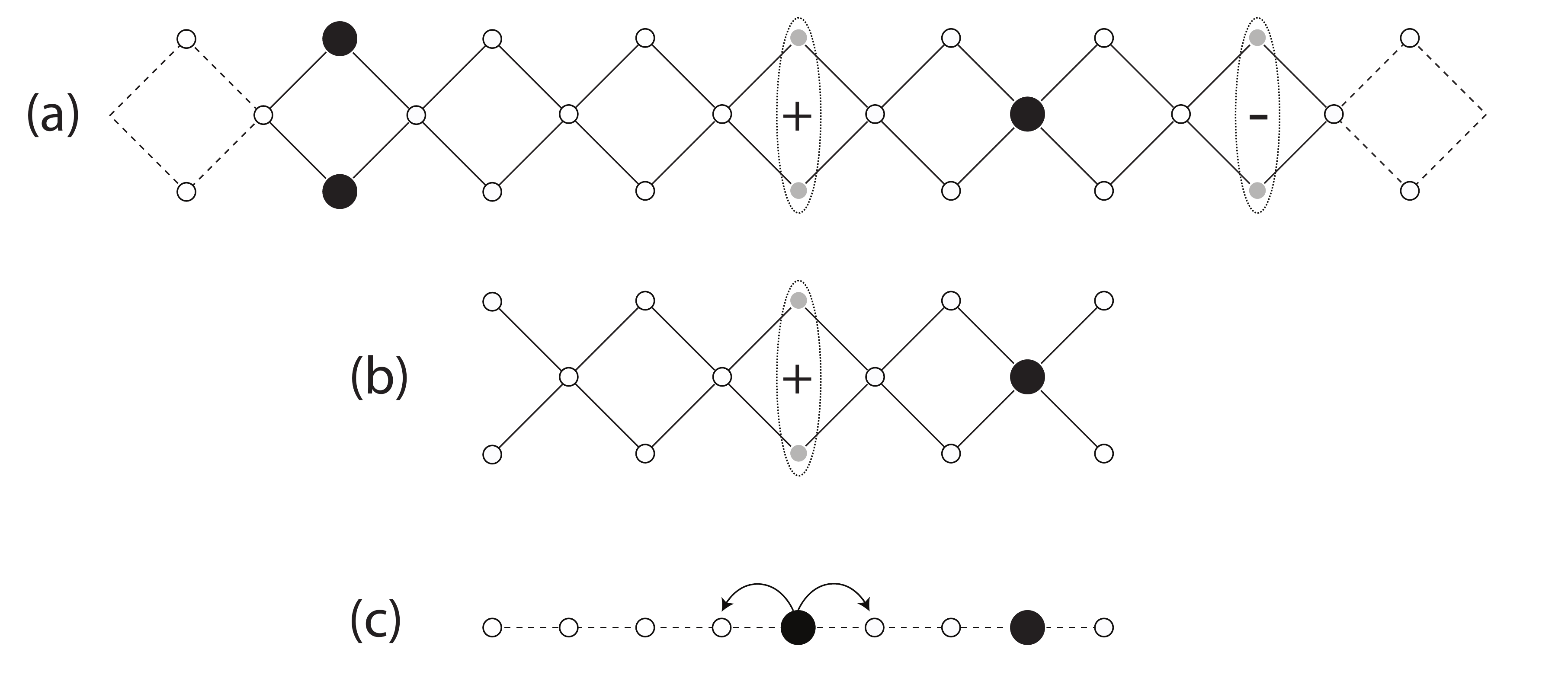}
    \caption{(a) State with localized as well as itinerant particles; (b) Open boundary region where the itinerant fermions in the state given in (a) may move. Note that the open boundary regions  terminate always at B and C sites. (c) Corresponding state of the linear lattice.}
    \label{fig:statesOB}
\end{figure}

Let us consider now the case of $N$ particles in a AB$_2$ chain with  $N_c$ unit cells.
Let us define the operators $B_{+,i}^{\dagger}=(B_i^{\dagger}+C_i^{\dagger})/\sqrt{2}$ and $B_{-,i}^{\dagger}=(B_i^{\dagger}-C_i^{\dagger})/\sqrt{2}$. Note that the product of these two operators creates two particles, one at site B$_i$ and the other at site C$_i$ as expected.
Using this new basis, the Hamiltonian can be rewritten in a simpler form
\begin{eqnarray}
    P_0 H P_0 = 
    -\sqrt{2} t \sum_{j=1}^{N_c}   
    && 
    \e{\I \phi_o/2N_c}  
    \left( 
    \prod_{ i \; \epsilon \;\mathcal{P}_j} 
    n^{h}_{i}   \times 
    A_j^\dagger B_{+,j+1}  
    \right. \label{eq:shortprojectedH}\\
     && \left. +  
    \prod_{ i \; \epsilon \;\mathcal{P}_{j+1}} 
    n^{h}_{i}   \times 
   B_{+,j+1}^\dagger A_{j+2}
    \right) 
    + \text{H.c.} \nonumber
\end{eqnarray}
One could be tempted to say that the localized states play no role in this simplified Hamiltonian since only the $B_{+,j}^\dagger$ operator appears, but that would be incorrect. 
A basis for the ground state subspace ($l=0$) can be constructed from states of the form
\begin{equation}
    \vert \Psi \rangle =\prod_{i=1}^{N_{BC}} B_{\alpha_i}^\dagger C_{\alpha_i}^\dagger \prod_{j=1}^{N_{B_-}} B_{-,\beta_j}^\dagger \prod_{n=1}^{N_{B_+}} B_{+,\gamma_n}^\dagger \prod_{m=1}^{N_{A}} A_{\mu_m}^\dagger
    \vert 0 \rangle
\end{equation}
where the sets $\{ \alpha \}$, $\{ \beta \}$, $\{ \gamma \}$, $\{ \mu \}$, $\{ \alpha \pm 1 \}$, $\{ \beta \pm 1 \}$, $\{ \gamma \pm 1 \}$ and $\{ \mu \pm 1 \}$   have no common elements in order to satisfy the $l=0$ condition and  $N_{BC}$, $N_{B_-}$, $N_{B_+}$ and $N_A$, are respectively the number  of localized pairs of fermions at sites B and C, of antibonding localized fermions, of itinerant bonding fermions  and of itinerant fermions at A sites, so that $2 N_{BC}+ N_{B_-} + N_{B_+} + N_A=N$.
Due to the projection condition, the localized fermions created by the $B_{-,j}^\dagger$ and $B_{+,j}^\dagger B_{-,j}^\dagger$ operators  act creating open boundaries for the hoppings of fermions. The Hamiltonian mixes states with different  configuration for the $A_j^\dagger$ and $B_{+,j}^\dagger$ products but the part relative to the localized fermions remains always the same.  In Fig.~\ref{fig:statesOB}, we illustrate this.

The diagonalization of the Hamiltonian given by Eq.~\eqref{eq:shortprojectedH} is achieved by first distributing the localized fermions (if any) and then solving the remaining problem for the itinerant fermions which move in regions confined by the localized particles, leading to a factorized form of the eigenstates, $\vert \text{itinerant fermions} \rangle \otimes \vert \text{localized fermions} \rangle $.
The eigenvectors and eigenvalues of the 1D t-V model  for twisted boundary conditions as well as open-boundary conditions can be obtained by use of the Bethe ansatz, but an equivalent but simpler algebraic solution is known in the strong-coupling limit.
In the following, we shortly review the main results of the algebraic solution. 

\subsection{Itinerant states}
Let us consider first the case when no particle is localized. In this case, the model becomes equivalent to the t-V chain apart from a renormalization of the hopping constant ($ t \rightarrow \sqrt{2} t$). The solution of the t-V chain relies in recognizing that the condition $l=0$ leads to a conservation of the number of links $N_{hp}$ and $N_{hh}$ so that the tight-binding term only exchanges the position of these links \cite{Dias2000}. Interpreting the (hp) links as non-interacting particles hopping in a chain where the empty sites are the (hh) links, the solution is attained.
Since the total number
of $(hh)$ and $(hp)$ links is $\tilde{L}=L-N$, the effective chain is reduced in relation to the real t-V chain. Note that $L$ is the number of sites in Fig.~\ref{fig:statesOB}(c), after the mapping to a 1D chain.
The fact that the tight-binding particles occupy two sites of the real t-V chain leads to a twisted boundary condition which is dependent on the momenta of the tight-binding particles and the  eigenvalues are given by \cite{Dias2000}
\begin{equation}
        E(\{\tilde{k}\},P)=-2 \sqrt{2} t \sum_{i=1}^{N}
        \cos \left(\tilde{k}_i-\dfrac{P}{\tilde{L}}  -{\phi \over L} \right)
        \label{eq:sevE}
\end{equation}
with $\tilde{k}=\tilde{n} \cdot 2\pi /\tilde{L}$,
and $P=n \cdot 2\pi /L$, with $\tilde{n}=0, \dots, \tilde{L}-1$
and $n=0, \dots, L-1$.
The set of  pseudo-momenta
$\{\tilde{k}\}$ and $P$  must satisfy the  following condition  
\begin{equation}
        P {L \over \tilde{L}}=\sum_{i=1}^{N} \tilde{k}
        \quad (\text{mod} \,\, 2\pi).
\end{equation}
The mapping of this solution of the t-V chain (with even number of sites) into the AB$_2$ chain without localized particles is direct,
with odd sites corresponding to A sites and even sites to sites B and C (which will be unoccupied or in a bonding configuration).

\subsection{Localized states}
Let us consider now a state where localized particles are present. Then one has one or more open boundary regions that terminate always in B and C sites as shown in Fig.~\ref{fig:statesOB}, i.e., the number of sites is odd (the B$_i$ and C$_i$ sites count as one site).
The solution for the itinerant particles in one open-boundary region is rather simple \cite{G'omez-Santos1993}. Again, a mapping to a system of free fermions in a reduced linear lattice is possible in a similar way, thinking of $(hp)$ links hopping in a background of $(hh)$ links, but it is simpler to state as in \cite{G'omez-Santos1993,G'omez-Santos1992} that the positions $\tilde{i}$ of particles in the reduced chain are given by the relation $\tilde{i}=i- N_i$ where $i$ is the position of the particle in the t-V 1D lattice and $N_i$ is the total number of particles between the initial site and site $i$. The number of sites of the reduced chain is  $\tilde{L}=L-N+1$.
The energy contribution of these itinerant fermions is $E_a=-2 \sqrt{2}t \sum_{i=1}^{N_a}\cos(k_i)$ with $k= \pi n/(N_{\text{red}}+1)$ and $n=1,2, \ldots,N_{\text{red}}$. Note that in the same state, and according to the distribution of the localized fermions, several confined regions may be present.
\subsection{Ground state}
The  contribution of localized fermions to the energy of a given state is zero, but that does not necessarily imply that the ground state will always correspond  to  the minimum number of localized fermions. As an example, let us consider $\rho=1/3$. An eigenstate for this filling is 
$ 
\prod_{m \text{ odd}}^{2N_c - 1} A_{m}^\dagger
\vert 0\rangle.
$
The respective energy is zero since this state corresponds to having the itinerant fermion band completely filled, so that the positive kinetic energies balance the negative kinetic energies. In this case, it is possible to construct lower energy eigenstates with localized fermion pairs (created by the operator $B_i^\dagger C_i^\dagger$), which allow room for the itinerant fermions to move and therefore lower the total kinetic energy.

This competition between itinerant and localized states will in fact start to occur at lower filling, $\rho=2/9$, since at this filling, the reduced chain is half-filled and therefore positive kinetic energy states will start to be filled when adding additional itinerant particles.
In order to lower the energy, one wants to keep the number of itinerant fermions just below half-filling and to have the maximum possible length for the reduced lattices where the itinerant particles move. Noting that a localized pair forbids the presence of particles in the two neighboring A sites, one concludes in order to maximize the number of sites available for the itinerant fermions these localized pairs should gather in a single cluster.

The ground-state energy is obtained from Eq.~\eqref{eq:sevE} for fillings less than 2/9. If $N$
is odd, all single-particle states with pseudo-momentum
$\tilde{k}$ between $\pm 2\pi/\tilde{L} \cdot (N-1)/2$ are
occupied and $\sum \tilde{k}=0$. Therefore,

\begin{equation}
        E_{\text{GS}}^{\text{odd}} 
        = -2 \sqrt{2} t 
        \dfrac{\sin 
        \left(
        \frac{\pi N}{2 N_c-N}
        \right)}{\sin 
        \left(
        \frac{\pi}{2 N_c-N}
        \right)
        }
        \cos 
        \left(
        \frac{\phi}{2 N_c}
        \right)
        \label{eq:EGS_odd}
\end{equation}

If $N$ is even, all states with
$\tilde{k}$ between $- 2\pi/\tilde{L} \cdot (N-2)/2$ and
$ 2\pi/\tilde{L} \cdot N/2$ or between $- 2\pi/\tilde{L} \cdot N/2$
and $ 2\pi/\tilde{L} \cdot (N-2)/2$
are occupied and $\sum \tilde{k}=\pm \pi/\tilde{L} \cdot N/L$. So,

\begin{equation}
        E_{\text{GS}}^{\text{even}} 
        = -2 \sqrt{2} t 
        \dfrac{\sin 
        \left(
        \frac{\pi N}{2 N_c-N}
        \right)}{\sin 
        \left(
        \frac{\pi}{2 N_c-N}
        \right)
        }
        \cos 
        \left(
        \frac{\pi-\phi}{2 N_c}
        \right)
        \label{eq:EGS_even}
\end{equation}
In the thermodynamic limit, the difference in the last two expressions becomes irrelevant and one can write
\begin{equation}
        \dfrac{E_{\text{GS}}^{\text{itin}} }{N_c}
        = - 
        \dfrac{2 \sqrt{2} t 
        }{\pi}
        (2 -3 \rho)
        \sin 
        \left(
        \frac{\pi \rho}{\frac{2}{3} -\rho}
        \right)
\end{equation}

For filling larger than 2/9, it is energetically favorable to have localized pairs of fermions in consecutive unit cells so that only one region exists for itinerant fermions, with $2N_c-2N_{BC}-1$ sites and the respective reduced lattice will have $\tilde{L}=2N_c-N$ sites.  The number of itinerant fermions is $N_{\text{itin}}=N-2 N_{BC}$. The number of BC localized pairs is such that the band for the itinerant fermions in the respective reduced lattice is as near as possible from half-filling, so that $N_{BC}=\text{Int}(3N/4-N_c/2)+1$ or $N_{BC}=\text{Int}(3N/4-N_c/2)$. 
In the thermodynamic limit, one can write $N_{\text{itin}}=N_c-N/2$ and one has 
a closely packed localized  cluster with length equal to $N_{BC}/2=3N/8-N_c/4$.
\begin{figure}[t]
	\begin{center}
		$\begin{array}{c}
		\includegraphics[width=7.3cm]{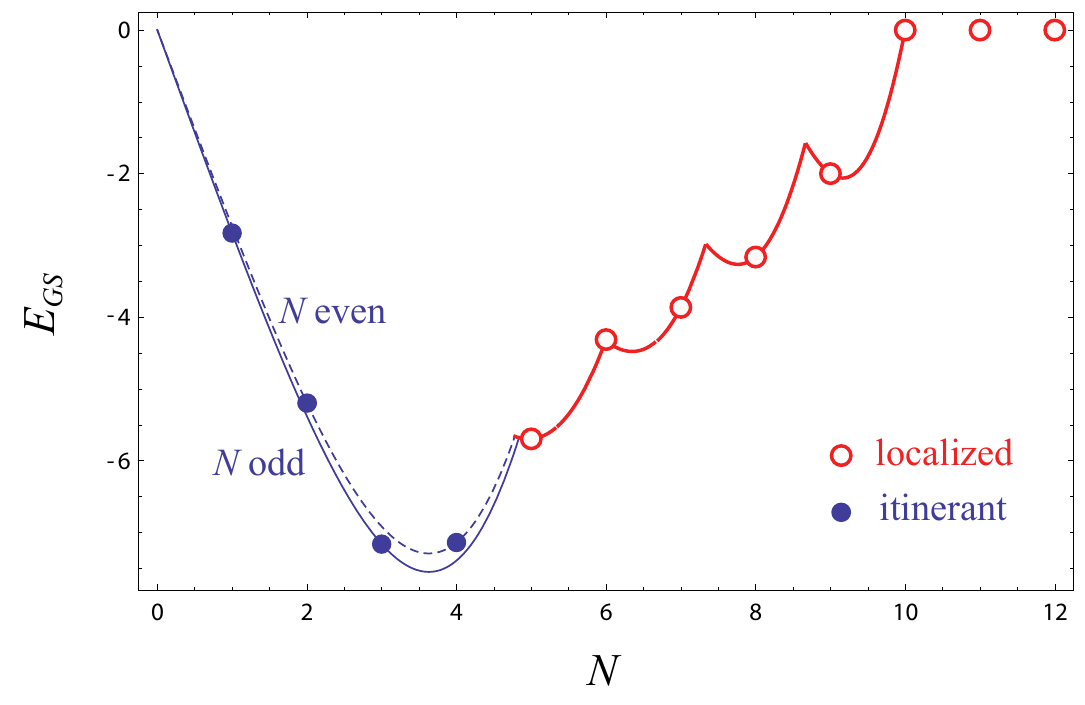}  
		\\		
		(a) 
		\\
		\\
		\includegraphics[width=7.5cm]{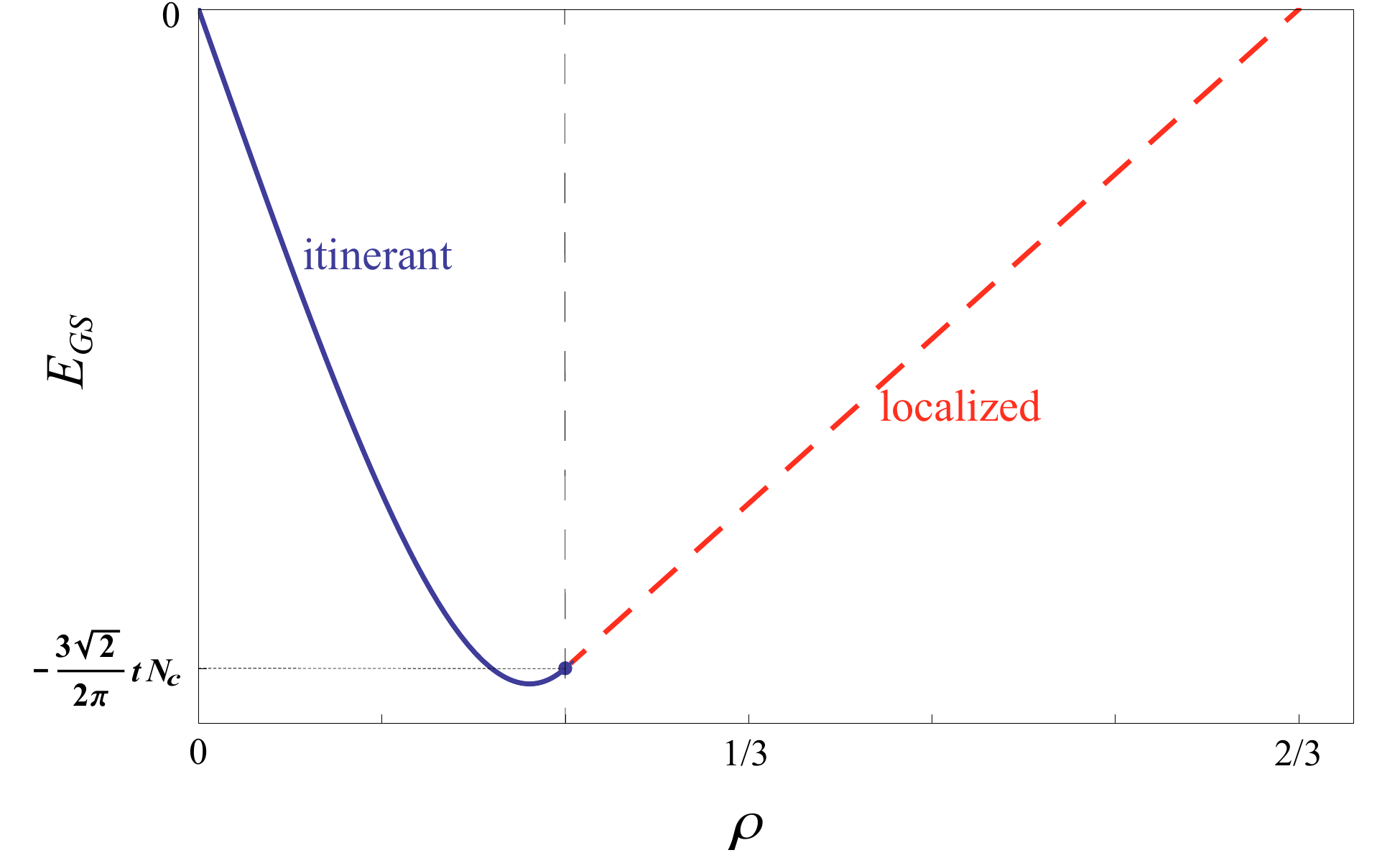}  
		\\
	(b)
		\end{array}$
	\end{center}
    \caption{(a) Ground state energy as a function of filling of the AB$_2$ chain in the strong-coupling limit $V=\infty$ and with $N_c=6$. The curves are the analytical results given by Eqs.~\eqref{eq:EGS_odd}, \eqref{eq:EGS_even} and \eqref{eq:EGS_OB} and the dots are the energy levels obtained by numerical diagonalization of the $V=\infty$ AB$_2$ chain. (b) Ground state energy in the thermodynamic limit as a function of filling in the strong-coupling limit $V=\infty$. The transition between a metallic and an insulating ground state occurs exactly at $\rho=2/9$ and the minimum energy is obtained when $\rho \approx 0.2$. }
    \label{fig:Nc6}
\end{figure}

The ground-state energy in this case is 
\begin{equation}
        E_{\text{GS}}^{\text{loc}} 
        = -2 \sqrt{2} t 
        \cos 
        \left[\frac{\pi}{2} \cdot
        \frac{N_{\text{itin}}+1}{2 N_c-N+1}
        \right] 
        \dfrac{
        \sin 
        \left[\frac{\pi}{2} \cdot
        \frac{N_{\text{itin}}}{2 N_c-N+1}
        \right] 
        }{
        \sin 
        \left[\frac{\pi}{2} \cdot
        \frac{1}{2 N_c-N+1}
        \right]
        }
        \label{eq:EGS_OB}
\end{equation}
which for a large AB$_2$ chain simplifies to 
\begin{equation}
        \dfrac{E_{\text{GS}}^{\text{loc}} }{N_c}
        = - 
        \dfrac{2 \sqrt{2} t 
        }{\pi}
        (2 -3 \rho),
\end{equation}
where $\rho=N/3N_c$. Since this ground state is localized, no magnetic flux dependence is present.
Note that for $N=2N_c-1$ the reduced lattice has only one site and the only possible energy is zero. Also for $N=2N_c$, one has a Wigner crystal like state with $N_{BC}=N_c$ and zero energy. Also for $N=2N_c-2$, the reduced lattice has two sites and and the number of itinerant fermions is zero or two, therefore the only possible energy is zero.

The density in the compact cluster of BC localized pairs present in the ground state for $\rho \geq 2/9$ has density equal to 2/3, while the  region available for the movement of the itinerant fermions has constant density $N_{\text{itin}}/(\frac{9}{2}  N_{\text{itin}}-1)$ which for large $N_{\text{itin}}$ is approximately constant and equal to 2/9. So as one increases the filling in the AB$_2$ chain, one is reducing the length of a phase with lower density and increasing the size of the higher density phase.

Phase separation  has been observed in other 1D and 2D Hamiltonians \cite{Emery1990,Ogata1991}.
In the 1D t-J model, phase separation occurs for $ J/t=2.5-3.5$, with the system divided into an electron-rich region and a hole-rich region \cite{Ogata1991}. An open-boundary  Heisenberg model rules the dynamic of  spin degrees of freedom of the electron-rich phase. 
Particularly relevant to our study of the AB$_2$ chain is the phase separation of the anisotropic  Heisenberg model or XXZ model, which can be mapped onto the 1D t-V model.  
In this case, phase separation corresponds to the appearance of the ferromagnetic phase in the XXZ model \cite{Haldane1980}.
The anisotropy constant of the XXZ model, $\Delta = J_\parallel /J_\perp$ under the Jordan-Wigner transformation becomes $\Delta = -V/2 \vert t \vert$ and for both models the phase separation occurs for $\Delta \geq 1$ \cite{Haldane1980,Nakamura1997}, that is, phase separation occurs for attractive nearest-neighbor interaction (ferromagnetic interaction).
In the t-V AB$_2$ chain, we have shown that phase separation occurs for strong \textit{repulsive} interaction. Note that a spinless fermion in the t-V model corresponds to an up-spin in XXZ model while a hole corresponds to a down-spin, and therefore the lower and higher density phases of the AB$_2$ chain correspond to different magnetization regions of the \textit{antiferromagnetic}  XXZ AB$_2$ model.

\subsection{Luttinger liquid description}

One of the most interesting points concerning the t-V AB$_2$ chain is the following: is the  t-V AB$_2$ chain a Luttinger liquid?
That is, can the low energy excitations of this model be described as bosonic charge density fluctuations governed by the harmonic Hamiltonian \cite{Haldane1980,*Haldane1981,*Haldane1981a} 
\begin{equation}
	H_{LL}=v_S \sum_q \vert q \vert b_q^\dagger b_q + \frac{\pi}{2L}[v_N (N-N_o)^2+v_J J^2] ?
\end{equation}
In the LL Hamiltonian,  $N$, $J$, $L$, $v_N$, $v_J$, and $v_S$ are, respectively, the particle number, current number, system length, particle velocity, current velocity, and sound-wave velocity.
One is easily tempted to calculate these Luttinger liquid parameters from the previous strong coupling results,  obtaining for large $L$ and for $\rho <2/9$, 
\begin{equation}
    v_N={1\over \pi} {\partial^2 (E_T/L) \over \partial (N/L)^2}
    =\frac{36 \sqrt{2} t }{(2-3 \rho )^3} \sin \left(\frac{3 \pi  \rho }{2-3 \rho}\right)
\end{equation}
\begin{equation}
    v_J={\pi } {\partial^2 (E_T/L) \over \partial (\phi/L)^2}
    =\frac{9 (2-3 \rho ) t }{2\sqrt{2}} \sin \left(\frac{3 \pi  \rho }{2-3 \rho}\right)
\end{equation}
\begin{equation}
    v_S=\sqrt{v_N v_J}=\frac{9 \sqrt{2} t }{2-3 \rho } \sin \left(\frac{3 \pi  \rho }{2-3 \rho}\right)
\end{equation}
where $\rho={N \over L}$ and $v_N$, $v_J$, and $v_S$ are respectively
the particle, current, and sound velocities. Note that $N/L=3 \rho /2$ since $L=2 N_c$.
These are  results similar to those obtained for the strong coupling  t-V  ring\cite{G'omez-Santos1993,G'omez-Santos1992} in the thermodynamic limit. The Luttinger liquid parameter 
\begin{equation}
    \frac{1}{K_\rho}= e^{-2 \cal{\psi}}=v_N/ v_S=\frac{4}{(2-3 \rho )^2}
\end{equation}
determines the 
anomalous correlation exponents and is  filling dependent  reflecting the reduction of the effective size of the chain with filling.

For filling larger than 2/9,  the ground state  has the localized pairs of fermions in a compact cluster so that only one region exists for itinerant fermions, with $2N_c-2N_{BC}-1$ sites and the respective reduced lattice will have $\tilde{L}=2N_c-N$ sites. One again is tempted to describe the low energy behavior of this system of itinerant fermions in this open-boundary lattice using as above a LL description. 
With open boundaries, one loses the  translation invariance but it has been shown by several authors  that an open-boundary bosonization can still be carried out \cite{Fabrizio1995,Eggert1996,Wang1996,Voit2000,Meden2000}. Many of these studies of LLs with open boundaries were motivated by the fact that the introduction of a single local impurity in a Luttinger liquid breaks the 1D system in two parts, at least in what concerns the low energy properties of the system \cite{Kane1992,Kane1992a}.
There is a clear analogy between the role of these local impurities and the localized fermions in the AB$_2$ chain. For  very large  $L$, one expects  the solution of the system to become  less dependent on the boundary conditions and therefore the LL velocities  are the same as those of an equivalent system  with periodic boundary conditions {\it at the same filling and with the same lattice size} (since the LL exponents are dependent on filling).
But we have seen that the filling in this open-boundary region is constant and equal to 2/9, and since the LL velocities depend only on filling, one obtains 
\begin{equation}
    v_N=\frac{243 t}{8 \sqrt{2}}, \quad
    v_J=3 \sqrt{2} t, \quad
    v_S=\frac{27 t}{2\sqrt{2}}, \quad
    e^{-2 \cal{\psi}}=\frac{9}{4}.
\end{equation}
Note that this ground state is $N_c$ degenerate due to the translation invariance of the phase separation boundaries.

The missing point in the previous analysis is that the LL Hamiltonian does not describe the excitations of ground state that involve the creation of an additional localized anti-bonding fermion or an additional localized BC pair of fermions.
However this is only important if these excitations are low energy states. In a qualitative picture, the strong coupling AB$_2$ chain can be described as if there were a flat band at zero energy with $N_c$ two-level sites. This is similar to the noninteracting case, but now the number of localized fermions in such flat band is twice as many. 
For $\rho \ll 2/9$,  the creation of a localized anti-bonding fermion or a localized BC pair of fermions implies an excitation energy of the order of $t$ and for energies and temperatures much less than this value, the LL description  is valid. 
As the filling approaches 2/9, these excitations become low lying (since the Fermi level goes to zero) and must clearly be taken into account in the low temperature description of the system.
In this situation, even at very low temperature,  the low energy AB$_2$ set of  eigenvalues becomes a complex  mix of the sets of eigenvalues of LLs with different sizes, fillings,  boundary conditions, and  LL velocities.
Despite these remarks, one should note that the strong coupling LL velocities do 
indeed characterize the correlations that do not involve the creation of additional anti-bonding states or  additional localized BC pairs of fermions. For example, a two-point ground-state  correlation involving only A sites such as the Green's function restricted to  A sites.

For $\rho > 2/9$, the compressibility 
\begin{equation}
	\frac{1}{\kappa}=\frac{1}{L}\frac{\partial^ 2 E_0(\rho)}{\partial \rho^ 2}
\end{equation} 
is infinite due to the linear behavior of the ground-state energy. This is the expected behavior of a phase separated ground state and is known to occur also in the t-J model as well as in the XXZ model \cite{Nakamura1997}. In these models, the compressibility diverges as $(\Delta_c -\Delta)^{-1}$ as  one approaches the phase separation critical value (which is $\Delta_c=1$ in the case of the attractive t-V chain) by increasing  the interaction constant \cite{Nakamura1997}.
The compressibility can also be calculated from the LL relation
\begin{equation}
	\kappa =\frac{2K_\rho}{\pi v_s}.
\end{equation}
In the XXZ model (or in the equivalent attractive t-V model), the compressibility diverges since the sound velocity vanishes and the Luttinger parameter $K_\rho$ diverges as $\Delta \rightarrow \Delta_c$.
From the last two relations, one easily concludes that as the filling in the AB$_2$ approaches 2/9, the compressibility does not diverge, since the curvature of the plot of $E(\rho)$ in Fig.~\ref{fig:Nc6}(b) does not vanish just below $\rho=2/9$ or equivalently,  neither the sound velocity vanishes nor the Luttinger parameter $K_\rho$ diverges as $\rho \rightarrow 2/9$.

For $\rho < 2/9$, the current velocity gives the charge stiffness at zero temperature \cite{Kohn1964,Shastry1990}
\begin{equation}
       D_c={1 \over 2} 
       {\partial^2(E_T/L) \over \partial(\phi/L)^2} \mid_{\phi_c=0}.
\end{equation}
For $\rho \geq 2/9$,  the charge stiffness is zero reflecting the open boundaries condition for the itinerant fermions and consequent zero dependence of the energy levels on magnetic flux. This discontinuity of the Drude weight 
at the phase boundary is known to occur in other 1D models with phase separation \cite{Nakamura1997}.
In this case, the current velocity should correspond to a very small frequency peak in the optical conductivity since it is known that the open-boundary condition shifts the spectral weight associated to the Drude peak to a finite frequency peak \cite{Fye1991}. 

\section{Implications for the extended Hubbard AB$_2$ model}

In this section, we discuss the relevance of the results obtained in this paper for the t-V AB$_2$ model for the spinful extended Hubbard model in the AB$_2$ geometry and in the strong-coupling limit $U\gg t$ and $U \gg V$. 

First, let us recall the known facts about the strong coupling Hubbard model in a ring \cite{Ogata1990,Schadschneider1995,Gebhard1997,Dias1992,Peres2000} and in a 1D chain with open-boundary conditions \cite{Arrachea1994}.
The  extended Hubbard Hamiltonian for a ring with
$L$ sites is given by
\begin{eqnarray}
     H &=& -t \sum_{i} (c_{i\sigma}^\dagger c_{i+1\sigma}
     + c_{i+1\sigma}^\dagger c_{i\sigma})  \nonumber \\
     &&+ U \sum_i n_{i\uparrow} n_{i\downarrow} + V \sum_i n_{i} n_{i+1},
     \label{eq:H1}
\end{eqnarray}
where the creation (annihilation) of an electron at site $i$ with
spin $\sigma$ is denoted by $c^\dagger_{i \sigma}$ ($c_{i \sigma}$)
with $n_{i \sigma}$ being the number operator $n_{i \sigma}=
c^\dagger_{i \sigma} c_{i \sigma}$ and $n_i=n_{i \uparrow}+n_{i \downarrow}$.
When $t=0$, all states
with the same number $N_d$ of doubly occupied sites and the same number
$\sum_{\sigma \sigma'} N_{\sigma \sigma'}$
of nearest-neighbor occupied sites  are degenerate,
where $N_{\sigma \sigma'}$ is the total number of nearest-neighbor pairs
with spin configuration $\sigma \sigma'$.
The eigenvalues of the extended Hubbard model in the atomic limit
are given by $E (N_d,\{N_{\sigma \sigma'}\})=N_d \cdot U+
\sum_{\sigma \sigma'} N_{\sigma \sigma'} \cdot V$.
Here we will only address the low energy subspace of the strong-coupling limit where $N_d=0$ to make this discussion simpler.

If we consider the Hubbard ring with $t \ll U$, but    $V=0$, one has the so-called Harris-Lange model \cite{Harris1967} and the model eigenfunctions can be written as a tensorial product of the eigenfunctions of a tight-binding model of independent spinless fermions in the ring (where the spinless fermions are the electrons deprived of spin) and the eigenfunctions of an Heisenberg model (with exchange constant 
$J=t^2/U$) for the  spins of the electrons in a reduced chain \cite{Ogata1990,Schadschneider1995,Gebhard1997,Dias1992,Peres2000}.
The spinless fermions  ring is threaded by a
fictitious magnetic flux $\phi=q_s$
generated by the spin 
configurations in the reduced Heisenberg chain (where $q_s$ is the total spin momentum)
and the eigenvalues to order $t$ are given by
\begin{equation}
        E(k_1,\ldots ,k_{N_h+N_d})=2t \sum_{i=1}^{N_h}
        \cos \left(k_i-{q_s\over L}\right),
        \label{eq:eigen}
\end{equation}
where $k_i=(2\pi/L) n_i$, $n_i=0,\ldots,L-1$ are the momenta of the holes in the spinless ring.

The  nearest-neighbor interaction is obviously independent of spin of the electrons that occupy nearest-neighbor sites and it is easily introduced in the previous picture so that for  $U\gg t$ and $U \gg V$, the $N_d=0$ spectrum of the Hubbard model to the order of $t$ is that of interacting spinless fermions in a ring threaded by a
fictitious magnetic flux $\phi=q_s$ (since the nearest-neighbor repulsion between
electrons leads to an nearest-neighbor repulsion between spinless fermions),
that is, the  extended Hubbard model in a ring in the limit $U \rightarrow \infty$ has the energy dispersion of the spinless t-V chain. The only effect of spin in this limit is the generation of a fictitious flux and the increase of degeneracy.

If now we consider the strong coupling Hubbard chain with open-boundary conditions, the solution is even simpler. As stated in Ref. \cite{Arrachea1994}, for $\rho <1$, the physics of the model is the same as that of a spinless tight-binding model. In fact, the electrons hop along the chain but the spin configuration of the electrons remains always the same and becomes irrelevant for the determination of the eigenvalues of the model. The different spin configurations only contribute to the degeneracy of the energy levels. The same reasoning as above can be followed and one concludes that the extended Hubbard chain with open-boundary conditions  and in the limit $U\gg t$ and $U \gg V$ has the same energy spectrum as the t-V chain with open-boundary conditions.

Let us now consider the extended Hubbard model in the AB$_2$ geometry,
\begin{eqnarray}
    H &=& H_0 + V \sum_j \left(n^A_j + n^A_{j+1}\right)\left( n^B_j + n^C_j \right)  \nonumber \\
    &+& U \sum_i (n^A_{i\uparrow} n^A_{i\downarrow} 
    +n^B_{i\uparrow} n^B_{i\downarrow}+n^C_{i\uparrow} n^C_{i\downarrow}),
\end{eqnarray}
where 
\begin{eqnarray}
    H_0 = -t \sum_{\sigma} \sum_{j=1}^{N_c} &  & \left[ \e{\I \phi_o/2N_c} ( A_{j\sigma}^\dagger B_{j\sigma}  + B_{j\sigma}^\dagger A_{j+1\sigma}) \right. \\
    & + & \left. \e{-\I \phi_i/2N_c} ( C_{j\sigma}^\dagger A_{j\sigma} + A_{j+1\sigma}^\dagger C_{j\sigma} ) \right] + \text{H.c.} \nonumber
\end{eqnarray}
We consider the limit $U \rightarrow \infty$ where no doubly occupied sites are possible.

In the AB$_2$ geometry, we can not state as we did for the Hubbard chain that for $V=0$ the model can be mapped into a system of spinless fermions in a tight-binding AB$_2$ chain threaded by  fictitious magnetic flux generated by the spin configuration momentum.
The reason is that in the case of the Hubbard chain, only  hoppings at the boundaries generate a different spin configuration and all the  different configurations generated  can be obtained from one of them applying a circular permutation operator \cite{Ogata1990,Schadschneider1995,Gebhard1997,Dias1992,Peres2000}.
 However, in the AB$_2$ geometry,  besides the circular permutation of the spin configuration due the hoppings at the boundaries, additional spin mixing can also be generated by electrons hopping  along one plaquette (which circularly permute a subset of two or three spins). However, this additional  exchange process requires the existence of an occupied  nearest-neighbor pair of sites in that plaquette as an intermediate step. 
Therefore, as $V$ is increased, this exchange process is inhibited.

Let us therefore address the extended Hubbard AB$_2$ model in the limit  $t \ll V\ll U$. In this case the additional plaquette exchange does not occur.
We show below that the low energy spectrum of this model, that is, the set of eigenvalues corresponding to eigenstates with  no doubly occupied sites and no
occupied  nearest-neighbor pair of sites  will be the same as that of the t-V AB$_2$ model in the limit $t \ll V$, but with enlarged degeneracy just like for the extended Hubbard ring in the strong-coupling limit. The picture behind this result will be a mix of the pictures presented for the Hubbard ring and for the Hubbard chain with open-boundary conditions.

First  we show that the itinerant and localized states of the extended Hubbard AB$_2$ model remain the same as those of the strong coupling t-V AB$_2$ model in what concerns the charge distribution, but now one has a considerable larger set of states due the spin degrees of freedom. 
We adopt the same basis as that of the previous section, but now spin must be considered, that is, the bonding and anti-bonding states have a spin degree of freedom.
In the previous section we have shown that two type of localized fermions could occur: i) one-particle localized states corresponding to a single particle in one plaquette in a standing wave state, which obviously remains localized if the particle has spin; ii) two-particle BC localized states induced by the large nearest-neighbor repulsion. The spinless localized pair maps into four localized BC pairs with different spin configurations which again remain localized due to the  large nearest-neighbor repulsion.  
Concerning the itinerant particles, the Hamiltonian does not alter the spin of the bonding states as the particles hop along the AB$_2$ chain.

Therefore, we can construct the eigenstates of the extended Hubbard AB$_2$ model in the limit case $t \ll V\ll U$ in an equivalent way to that followed in the previous section. 
If no localized particles are present, the extended Hubbard AB$_2$ model in the limit case $t \ll V\ll U$ can be mapped into  
the extended Hubbard ring in the same limit and the energy levels are  given by 
\begin{equation}
        E(\{\tilde{k}\},P)=-2 \sqrt{2} t \sum_{i=1}^{N}
        \cos \left(\tilde{k}_i-\dfrac{P}{\tilde{L}}  -{q_s+\phi \over L} \right)
\end{equation}
where $\tilde{k}_i$ and $\tilde{L}$ have the same definition as for the t-V ring in the strong-coupling limit, $q_s$  has the same definition as for the  Hubbard chain and $\phi$ is a external magnetic flux (but keeping a zero flux in each plaquette). In the thermodynamic limit, the effect of the fictitious and external fluxes becomes irrelevant.

If localized particles are present, they create open boundary regions where itinerant electrons (in bonding states when at sites B and C) will move according to the extended Hubbard AB$_2$ model in the strong-coupling limit. Each of these open boundary systems has a fixed number of particles and can be mapped as above  into the extended Hubbard chain in the same limit, but now with open-boundary conditions. In this case, as stated before, the spin configuration is irrelevant and contributes only to the degeneracy of the energy levels which are exactly those described in the previous section for the spinless t-V AB$_2$ model with open-boundary conditions.
Since the energy levels in the thermodynamic limit are the same as those of the t-V  AB$_2$ model in the strong-coupling limit, phase separation will also occur in the extended Hubbard AB$_2$ ring and precisely at the same filling.
This exact correspondence required the strong-coupling limit and was obtained  only for the low energy subspace.
Higher energy subspaces involve the presence of occupied nearest-neighbor pairs of sites or of double occupancies, but a similar approach can in principle be followed again with a mapping to higher energy subspaces of the extended Hubbard chain.

For intermediate values of the nearest-neighbor interaction $V$, the mapping is no longer exact, but qualitatively 
 the features described for the t-V AB$_2$ chain should be expected in the extended Hubbard AB$_2$ chain (with $U \gg t$ and $U \gg V$). For example we expect a similar evolution for the ground-state energy with decreasing $V$, but with different evolution of filling. Note that the ground-state energy for non-interacting electrons (with spin) as a function of filling has the same form of Fig.~\ref{fig:energyGS}, but the spin degeneracy of the single-particle eigenvalues implies that the filling values as well as the ground-state energy must be multiplied by two. This is equivalent to  distributing spinless particles among two independent AB$_2$ chains.
\section{Conclusion}

As mentioned in the introduction, the 1D anisotropic  Heisenberg model (the XXZ model) can be mapped using the Jordan-Wigner transformation into the 1D  spinless t-V model (with an additional ``chemical potential'' term, which can also be interpreted as an on-site energy). The ground state filling of the t-V model is related to the ground state magnetization of the XXZ model, which can be controlled by the application of an external magnetic field to the XXZ model. This external field does not change the eigenstates of the XXZ model, but creates an additional chemical potential term in the t-V model which allows to control the filling of the ground state. A ferromagnetic Heisenberg interaction means that the spinless fermions attract each other while an antiferromagnetic exchange implies a repulsive interaction between nearest-neighbor spinless fermions. A Jordan-Wigner transformation is also possible in quasi-1D or 2D lattices (see Ref. \cite{Derzhko2001} for a review), and in particular, the XXZ AB$_2$ model in the strongly anisotropic limit can be mapped into the strong coupling t-V AB$_2$ model with additional phase factors which are nonlocal (and therefore usually treated in a mean-field approach) \cite{Derzhko2001}. In this paper, we have not gone into the details of the transformation of the XXZ AB$_2$ model into the t-V AB$_2$ model (which would require the determination of the phase factors and the introduction of the on-site energy terms), but some conclusions can be drawn on general arguments. 
First, one-magnon localized states in the strongly anisotropic XXZ AB$_n$ model corresponding to standing waves in a t-V AB$_n$ array will exist since no additional phase factors appear when only one spinless fermion is present (only one spin flip).
Also, states with several spin flips corresponding to localized magnons in different plaquettes are also eigenstates
since the XX term of the XXZ model gives a zero contribution in the regions between the 
localized magnons (since all spins are aligned).
So the one-particle localized states in the t-V AB$_2$ model correspond to localized and independent magnons created in a ferromagnetic background, which have been observed in frustrated magnetic systems under high magnetic fields but below the saturation field \cite{Schulenburg2002}.
Second, independently of the phase factors, the two-particle localized BC particles will remain the same because the nearest-neighbor interaction obtained in the Jordan-Wigner transformation retains the same form as in the 1D case, and since these two particles  are completely localized, the phase factors are irrelevant. Furthermore, these localized particles create open boundary regions for the itinerant fermions independently of the phase factors. So, we expect a similar  behavior of the XXZ AB$_2$ model in the strongly anisotropic limit  to the one described in this paper for the strong coupling t-V AB$_2$ model. Such two-particle localized states in the strong coupling t-V AB$_2$ model correspond to localized pairs of magnons in the the XXZ AB$_2$ model in the strongly anisotropic limit.

The results presented in Sec. IV for single-particle states in AB$_n$ lattices may also be relevant to Josephson junction arrays (with the same geometry) both in the quantum limit and in the high capacitance (classical) limit.
It is known that a Josephson junction AB$_2$ chain with half a flux quantum per plaquette exhibits a highly degenerate classical ground state reflecting the completely flat energy bands of the AB$_2$ tight-binding model for this value of flux \cite{Benoit2002}. The tunneling of Cooper pairs between the different superconducting islands in an AB$_2$ geometry can also be described using a bosonic tight-binding AB$_2$ model.
Furthermore, in the high-capacitance limit, the charging energy due to the electrostatic interaction  within each (large) superconducting  island can be neglected and the  Hamiltonian becomes a classical XY model, 
$
     H=-J\sum_{\langle ij\rangle } \cos(\phi_{i}-\phi_{j}-A_{ij})\,,
$
where $\phi_{i}$ is the superconducting phase of the island $i$, $J$ is the Josephson coupling and $A_{ij}$ is the phase shift due to the presence of an external magnetic field, obtained from the integral of the vector potential along the path from $i$ to $j$.
This Hamiltonian can be mapped into a one-particle tight-binding model with the same geometry and under the same magnetic flux \cite{Choi1985}. In most geometries, the minimum energy phase configuration of the Josephson junction array will be obtained from the state of minimum energy of the tight-binding model (if this state is homogeneous). In the most general case, the Hamiltonian can be interpreted as the mean energy of a phase vector  and the stable phase configuration of the Josephson junctions array is obtained minimizing this energy and this minimization may imply  mixing  tight-binding states of different bands.

Let us now compare the exact results of the strong coupling AB$_2$ chain with the mean-field results.
Comparing Figs.~\ref{fig:EGS_v} and \ref{fig:Nc6} (b), one concludes that the mean-field results overestimate the interaction energy contribution for $\rho <2/3$. Basically, within the mean-field approach the low energy band increases its energy with increasing $V$, missing the fact that itinerant states are possible, which avoids the positive energy contribution of the nearest-neighbor interaction as found in the strong-coupling limit.
The mean-field results are qualitatively correct in what concerns the energy interval for the ground-state energy for $\rho<2/3$. In Fig.~\ref{fig:EGS_v}, one can see that the ground-state energy remains negative even for large $V$ in the density range $0<\rho<2/3$. The minimum of the ground-state energy as a function of filling  shifts continuously to  lower filling in contrast with the $V=\infty$ result where such minimum occurs at $\rho \approx 0.2$.
Also the large slope for $\rho >2/3$ agrees with the fact that for such fillings nearest-neighbor pairs are present and such states have infinite energy when $V=\infty$.

To conclude, in this manuscript, the spinless AB$_2$ chain with nearest-neighbor Coulomb interactions has been studied for any filling and taking into account magnetic flux. In the case of independent fermions, a simple construction of the localized states that generate the flat bands both in the presence and absence of flux has been found and generalized for 1D or 2D arrays of quantum rings.
The $V/t$ versus filling  phase diagram of the AB$_2$ chain was obtained using a mean-field approach. The dependence on filling of the mean-field ground-state energy  agrees qualitatively with the exact ground-state energy for  infinite $V$. 
The ground-state energy for infinite nearest-neighbor repulsion  has a quantum critical point at filling $2/9$ where a metal-insulator transition occurs.
This transition reflects the phase separation between a high density phase ($\rho=2/3$) and a low-density phase ($\rho=2/9$) that occurs at fillings larger than $2/9$.
Such phase separation occurs because the infinite nearest-neighbor repulsion leads to the appearance  of two-particle localized states (besides the one-particle localized states due to the topology of the AB$_2$ chain). These localized states  create open boundary regions for itinerant carriers and in order for these itinerant fermions to have only negative kinetic energy, phase separation becomes favorable.
At low filling, the low energy properties of t-V AB$_2$ chain can be described by the spinless Luttinger Hamiltonian, but  for filling near or larger than 2/9, the AB$_2$ set of eigenvalues becomes a complex  mix of the sets of eigenvalues of LLs with different fillings,  boundary conditions, and  LL velocities. 
If the itinerant fermions have spin, but a very strong on-site repulsion is present (that is, in the case of the extended Hubbard model in the strong-coupling limit $ U \gg V \gg t$), the energy-dispersion relation to the order of the hopping integral remains the same 
as that of the spinless AB$_2$ model in the presence of a flux and phase separation occurs at the same filling.


\subsection*{Acknowledgements}
This work was partially supported by a grant from Fundação para a Ciência e a Tecnologia (Portugal), co-financed by FSE/POPH.

\bibliography{prb}

\begin{thebibliography}{10}%
\makeatletter
\providecommand \@ifxundefined [1]{%
 \ifx #1\undefined \expandafter \@firstoftwo
 \else \expandafter \@secondoftwo
\fi
}%
\providecommand \@ifnum [1]{%
 \ifnum #1\expandafter \@firstoftwo
 \else \expandafter \@secondoftwo
\fi
}%
\providecommand \enquote [1]{``#1''}%
\providecommand \bibnamefont  [1]{#1}%
\providecommand \bibfnamefont [1]{#1}%
\providecommand \citenamefont [1]{#1}%
\providecommand\href[0]{\@sanitize\@href}%
\providecommand\@href[1]{\endgroup\@@startlink{#1}\endgroup\@@href}%
\providecommand\@@href[1]{#1\@@endlink}%
\providecommand \@sanitize [0]{\begingroup\catcode`\&12\catcode`\#12\relax}%
\@ifxundefined \pdfoutput {\@firstoftwo}{%
 \@ifnum{\z@=\pdfoutput}{\@firstoftwo}{\@secondoftwo}%
}{%
 \providecommand\@@startlink[1]{\leavevmode\special{html:<a href="#1">}}%
 \providecommand\@@endlink[0]{\special{html:</a>}}%
}{%
 \providecommand\@@startlink[1]{%
  \leavevmode
  \pdfstartlink
   attr{/Border[0 0 1 ]/H/I/C[0 1 1]}%
   user{/Subtype/Link/A<</Type/Action/S/URI/URI(#1)>>}%
  \relax
 }%
 \providecommand\@@endlink[0]{\pdfendlink}%
}%
\providecommand \url  [0]{\begingroup\@sanitize \@url }%
\providecommand \@url [1]{\endgroup\@href {#1}{\urlprefix}}%
\providecommand \urlprefix [0]{URL }%
\providecommand \Eprint[0]{\href }%
\@ifxundefined \urlstyle {%
  \providecommand \doi [1]{doi:\discretionary{}{}{}#1}%
}{%
  \providecommand \doi [0]{doi:\discretionary{}{}{}\begingroup
  \urlstyle{rm}\Url }%
}%
\providecommand \doibase [0]{http://dx.doi.org/}%
\providecommand \Doi[1]{\href{\doibase#1}}%
\providecommand \bibAnnote [3]{%
  \BibitemShut{#1}%
  \begin{quotation}\noindent
    \textsc{Key:}\ #2\\\textsc{Annotation:}\ #3%
  \end{quotation}%
}%
\providecommand \bibAnnoteFile [2]{%
  \IfFileExists{#2}{\bibAnnote {#1} {#2} {\input{#2}}}{}%
}%
\providecommand \typeout [0]{\immediate \write \m@ne }%
\providecommand \selectlanguage [0]{\@gobble}%
\providecommand \bibinfo [0]{\@secondoftwo}%
\providecommand \bibfield [0]{\@secondoftwo}%
\providecommand \translation [1]{[#1]}%
\providecommand \BibitemOpen[0]{}%
\providecommand \bibitemStop [0]{}%
\providecommand \bibitemNoStop [0]{.\EOS\space}%
\providecommand \EOS [0]{\spacefactor3000\relax}%
\providecommand \BibitemShut [1]{\csname bibitem#1\endcsname}%
\bibitem{Gul'acsi2007}%
  \BibitemOpen
  \bibfield{author}{%
  \bibinfo {author} {\bibfnamefont{Z.}~\bibnamefont{Gul\'acsi}}, \bibinfo
  {author} {\bibfnamefont{A.}~\bibnamefont{Kampf}},\ and\ \bibinfo {author}
  {\bibfnamefont{D.}~\bibnamefont{Vollhardt}},\ }%
  \bibfield{journal}{%
  \Doi{10.1103/PhysRevLett.99.026404}{\bibinfo {journal} {Phys. Rev. Lett.}}\
  }%
  \textbf{\bibinfo {volume} {99}},\ \bibinfo {pages} {026404} (\bibinfo {month}
  {Jul}\ \bibinfo {year} {2007})%
  \bibAnnoteFile{NoStop}{Gul'acsi2007}%
\bibitem{Kikuchi2005}%
  \BibitemOpen
  \bibfield{author}{%
  \bibinfo {author} {\bibfnamefont{H.}~\bibnamefont{Kikuchi}}, \bibinfo
  {author} {\bibfnamefont{Y.}~\bibnamefont{Fujii}}, \bibinfo {author}
  {\bibfnamefont{M.}~\bibnamefont{Chiba}}, \bibinfo {author}
  {\bibfnamefont{S.}~\bibnamefont{Mitsudo}}, \bibinfo {author}
  {\bibfnamefont{T.}~\bibnamefont{Idehara}}, \bibinfo {author}
  {\bibfnamefont{T.}~\bibnamefont{Tonegawa}}, \bibinfo {author}
  {\bibfnamefont{K.}~\bibnamefont{Okamoto}}, \bibinfo {author}
  {\bibfnamefont{T.}~\bibnamefont{Sakai}}, \bibinfo {author}
  {\bibfnamefont{T.}~\bibnamefont{Kuwai}},\ and\ \bibinfo {author}
  {\bibfnamefont{H.}~\bibnamefont{Ohta}},\ }%
  \bibfield{journal}{%
  \Doi{10.1103/PhysRevLett.94.227201}{\bibinfo {journal} {Phys. Rev. Lett.}}\
  }%
  \textbf{\bibinfo {volume} {94}},\ \bibinfo {pages} {227201} (\bibinfo {month}
  {Jun}\ \bibinfo {year} {2005})%
  \bibAnnoteFile{NoStop}{Kikuchi2005}%
\bibitem{Macedo1995}%
  \BibitemOpen
  \bibfield{author}{%
  \bibinfo {author} {\bibfnamefont{A.~M.~S.}\ \bibnamefont{Mac\^edo}}, \bibinfo
  {author} {\bibfnamefont{M.~C.}\ \bibnamefont{dos Santos}}, \bibinfo {author}
  {\bibfnamefont{M.~D.}\ \bibnamefont{Coutinho-Filho}},\ and\ \bibinfo {author}
  {\bibfnamefont{C.~A.}\ \bibnamefont{Mac\^edo}},\ }%
  \bibfield{journal}{%
  \Doi{10.1103/PhysRevLett.74.1851}{\bibinfo {journal} {Phys. Rev. Lett.}}\ }%
  \textbf{\bibinfo {volume} {74}},\ \bibinfo {pages} {1851} (\bibinfo {month}
  {Mar}\ \bibinfo {year} {1995})%
  \bibAnnoteFile{NoStop}{Macedo1995}%
\bibitem{Montenegro-Filho2006}%
  \BibitemOpen
  \bibfield{author}{%
  \bibinfo {author} {\bibfnamefont{R.~R.}\ \bibnamefont{Montenegro-Filho}}\
  and\ \bibinfo {author} {\bibfnamefont{M.~D.}\ \bibnamefont{Coutinho-Filho}},\
  }%
  \bibfield{journal}{%
  \Doi{10.1103/PhysRevB.74.125117}{\bibinfo {journal} {Phys. Rev. B}}\ }%
  \textbf{\bibinfo {volume} {74}},\ \bibinfo {pages} {125117} (\bibinfo {month}
  {Sep}\ \bibinfo {year} {2006})%
  \bibAnnoteFile{NoStop}{Montenegro-Filho2006}%
\bibitem{Vidal1998}%
  \BibitemOpen
  \bibfield{author}{%
  \bibinfo {author} {\bibfnamefont{J.}~\bibnamefont{Vidal}}, \bibinfo {author}
  {\bibfnamefont{R.}~\bibnamefont{Mosseri}},\ and\ \bibinfo {author}
  {\bibfnamefont{B.}~\bibnamefont{Dou\ifmmode~\mbox{\c{c}}\else
  \c{c}\fi{}ot}},\ }%
  \bibfield{journal}{%
  \Doi{10.1103/PhysRevLett.81.5888}{\bibinfo {journal} {Phys. Rev. Lett.}}\ }%
  \textbf{\bibinfo {volume} {81}},\ \bibinfo {pages} {5888} (\bibinfo {month}
  {Dec}\ \bibinfo {year} {1998})%
  \bibAnnoteFile{NoStop}{Vidal1998}%
\bibitem{Tamura2002}%
  \BibitemOpen
  \bibfield{author}{%
  \bibinfo {author} {\bibfnamefont{H.}~\bibnamefont{Tamura}}, \bibinfo {author}
  {\bibfnamefont{K.}~\bibnamefont{Shiraishi}}, \bibinfo {author}
  {\bibfnamefont{T.}~\bibnamefont{Kimura}},\ and\ \bibinfo {author}
  {\bibfnamefont{H.}~\bibnamefont{Takayanagi}},\ }%
  \bibfield{journal}{%
  \Doi{10.1103/PhysRevB.65.085324}{\bibinfo {journal} {Phys. Rev. B}}\ }%
  \textbf{\bibinfo {volume} {65}},\ \bibinfo {pages} {085324} (\bibinfo {month}
  {Feb}\ \bibinfo {year} {2002})%
  \bibAnnoteFile{NoStop}{Tamura2002}%
\bibitem{Tasaki1998}%
  \BibitemOpen
  \bibfield{author}{%
  \bibinfo {author} {\bibfnamefont{H.}~\bibnamefont{Tasaki}},\ }%
  \bibfield{journal}{%
  \Doi{10.1088/0953-8984/10/20/004}{\bibinfo {journal} {J. Phys.: Condens.
  Matter}}\ }%
  \textbf{\bibinfo {volume} {10}},\ \bibinfo {pages} {4353} (\bibinfo {year}
  {1998})%
  \bibAnnoteFile{NoStop}{Tasaki1998}%
\bibitem{Tasaki1998a}%
  \BibitemOpen
  \bibfield{author}{%
  \bibinfo {author} {\bibfnamefont{H.}~\bibnamefont{Tasaki}},\ }%
  \bibfield{journal}{%
  \Doi{10.1143/PTP.99.489}{\bibinfo {journal} {Prog. Theor. Phys.}}\ }%
  \textbf{\bibinfo {volume} {99}},\ \bibinfo {pages} {489} (\bibinfo {year}
  {1998})%
  \bibAnnoteFile{NoStop}{Tasaki1998a}%
\bibitem{Tasaki1998b}%
  \BibitemOpen
  \bibfield{author}{%
  \bibinfo {author} {\bibfnamefont{H.}~\bibnamefont{Tasaki}},\ }%
  \bibfield{journal}{%
  \bibinfo {journal} {J. Phys.: Condens. Matter}\ }%
  \textbf{\bibinfo {volume} {10}},\ \bibinfo {pages} {4353} (\bibinfo {year}
  {1998})%
  \bibAnnoteFile{NoStop}{Tasaki1998b}%
\bibitem{Tasaki1995}%
  \BibitemOpen
  \bibfield{author}{%
  \bibinfo {author} {\bibfnamefont{H.}~\bibnamefont{Tasaki}},\ }%
  \bibfield{journal}{%
  \Doi{10.1103/PhysRevLett.75.4678}{\bibinfo {journal} {Phys. Rev. Lett.}}\ }%
  \textbf{\bibinfo {volume} {75}},\ \bibinfo {pages} {4678} (\bibinfo {month}
  {Dec}\ \bibinfo {year} {1995})%
  \bibAnnoteFile{NoStop}{Tasaki1995}%
\bibitem{Tasaki1994}%
  \BibitemOpen
  \bibfield{author}{%
  \bibinfo {author} {\bibfnamefont{H.}~\bibnamefont{Tasaki}},\ }%
  \bibfield{journal}{%
  \Doi{10.1103/PhysRevLett.73.1158}{\bibinfo {journal} {Phys. Rev. Lett.}}\ }%
  \textbf{\bibinfo {volume} {73}},\ \bibinfo {pages} {1158} (\bibinfo {month}
  {Aug}\ \bibinfo {year} {1994})%
  \bibAnnoteFile{NoStop}{Tasaki1994}%
\bibitem{Tasaki1992}%
  \BibitemOpen
  \bibfield{author}{%
  \bibinfo {author} {\bibfnamefont{H.}~\bibnamefont{Tasaki}},\ }%
  \bibfield{journal}{%
  \Doi{10.1103/PhysRevLett.69.1608}{\bibinfo {journal} {Phys. Rev. Lett.}}\ }%
  \textbf{\bibinfo {volume} {69}},\ \bibinfo {pages} {1608} (\bibinfo {month}
  {Sep}\ \bibinfo {year} {1992})%
  \bibAnnoteFile{NoStop}{Tasaki1992}%
\bibitem{Mielke1999}%
  \BibitemOpen
  \bibfield{author}{%
  \bibinfo {author} {\bibfnamefont{A.}~\bibnamefont{Mielke}},\ }%
  \bibfield{journal}{%
  \Doi{10.1103/PhysRevLett.82.4312}{\bibinfo {journal} {Phys. Rev. Lett.}}\ }%
  \textbf{\bibinfo {volume} {82}},\ \bibinfo {pages} {4312} (\bibinfo {month}
  {May}\ \bibinfo {year} {1999})%
  \bibAnnoteFile{NoStop}{Mielke1999}%
\bibitem{Mielke1992}%
  \BibitemOpen
  \bibfield{author}{%
  \bibinfo {author} {\bibfnamefont{A.}~\bibnamefont{Mielke}},\ }%
  \bibfield{journal}{%
  \Doi{10.1088/0305-4470/25/16/011}{\bibinfo {journal} {J. Phys. A-Math.
  Gen.}}\ }%
  \textbf{\bibinfo {volume} {25}},\ \bibinfo {pages} {4335} (\bibinfo {year}
  {1992})%
  \bibAnnoteFile{NoStop}{Mielke1992}%
\bibitem{Mielke1991}%
  \BibitemOpen
  \bibfield{author}{%
  \bibinfo {author} {\bibfnamefont{A.}~\bibnamefont{Mielke}},\ }%
  \bibfield{journal}{%
  \Doi{10.1088/0305-4470/24/14/018}{\bibinfo {journal} {J. Phys. A-Math.
  Gen.}}\ }%
  \textbf{\bibinfo {volume} {24}},\ \bibinfo {pages} {3311} (\bibinfo {year}
  {1991})%
  \bibAnnoteFile{NoStop}{Mielke1991}%
\bibitem{Mielke1991a}%
  \BibitemOpen
  \bibfield{author}{%
  \bibinfo {author} {\bibfnamefont{A.}~\bibnamefont{Mielke}},\ }%
  \bibfield{journal}{%
  \Doi{10.1088/0305-4470/24/2/005}{\bibinfo {journal} {J. Phys. A-Math. Gen.}}\
  }%
  \textbf{\bibinfo {volume} {24}},\ \bibinfo {pages} {L73} (\bibinfo {year}
  {1991})%
  \bibAnnoteFile{NoStop}{Mielke1991a}%
\bibitem{Mielke1993}%
  \BibitemOpen
  \bibfield{author}{%
  \bibinfo {author} {\bibfnamefont{A.}~\bibnamefont{Mielke}}\ and\ \bibinfo
  {author} {\bibfnamefont{H.}~\bibnamefont{Tasaki}},\ }%
  \bibfield{journal}{%
  \Doi{10.1007/BF02108079}{\bibinfo {journal} {Commun. Math. Phys.}}\ }%
  \textbf{\bibinfo {volume} {158}},\ \bibinfo {pages} {341} (\bibinfo {year}
  {1993})%
  \bibAnnoteFile{NoStop}{Mielke1993}%
\bibitem{Derzhko2005}%
  \BibitemOpen
  \bibfield{author}{%
  \bibinfo {author} {\bibfnamefont{O.}~\bibnamefont{Derzhko}}\ and\ \bibinfo
  {author} {\bibfnamefont{J.}~\bibnamefont{Richter}},\ }%
  \bibfield{journal}{%
  \Doi{10.1103/PhysRevB.72.094437}{\bibinfo {journal} {Phys. Rev. B}}\ }%
  \textbf{\bibinfo {volume} {72}},\ \bibinfo {pages} {094437} (\bibinfo {month}
  {Sep}\ \bibinfo {year} {2005})%
  \bibAnnoteFile{NoStop}{Derzhko2005}%
\bibitem{Derzhko2004}%
  \BibitemOpen
  \bibfield{author}{%
  \bibinfo {author} {\bibfnamefont{O.}~\bibnamefont{Derzhko}}\ and\ \bibinfo
  {author} {\bibfnamefont{J.}~\bibnamefont{Richter}},\ }%
  \bibfield{journal}{%
  \Doi{10.1103/PhysRevB.70.104415}{\bibinfo {journal} {Phys. Rev. B}}\ }%
  \textbf{\bibinfo {volume} {70}},\ \bibinfo {pages} {104415} (\bibinfo {month}
  {Sep}\ \bibinfo {year} {2004})%
  \bibAnnoteFile{NoStop}{Derzhko2004}%
\bibitem{Derzhko2010}%
  \BibitemOpen
  \bibfield{author}{%
  \bibinfo {author} {\bibfnamefont{O.}~\bibnamefont{Derzhko}}, \bibinfo
  {author} {\bibfnamefont{J.}~\bibnamefont{Richter}}, \bibinfo {author}
  {\bibfnamefont{A.}~\bibnamefont{Honecker}}, \bibinfo {author}
  {\bibfnamefont{M.}~\bibnamefont{Maksymenko}},\ and\ \bibinfo {author}
  {\bibfnamefont{R.}~\bibnamefont{Moessner}},\ }%
  \bibfield{journal}{%
  \Doi{10.1103/PhysRevB.81.014421}{\bibinfo {journal} {Phys. Rev. B}}\ }%
  \textbf{\bibinfo {volume} {81}},\ \bibinfo {pages} {014421} (\bibinfo {month}
  {Jan}\ \bibinfo {year} {2010})%
  \bibAnnoteFile{NoStop}{Derzhko2010}%
\bibitem{Tanaka2007}%
  \BibitemOpen
  \bibfield{author}{%
  \bibinfo {author} {\bibfnamefont{A.}~\bibnamefont{Tanaka}}\ and\ \bibinfo
  {author} {\bibfnamefont{H.}~\bibnamefont{Tasaki}},\ }%
  \bibfield{journal}{%
  \Doi{10.1103/PhysRevLett.98.116402}{\bibinfo {journal} {Phys. Rev. Lett.}}\
  }%
  \textbf{\bibinfo {volume} {98}},\ \bibinfo {pages} {116402} (\bibinfo {month}
  {Mar}\ \bibinfo {year} {2007})%
  \bibAnnoteFile{NoStop}{Tanaka2007}%
\bibitem{Duan2001}%
  \BibitemOpen
  \bibfield{author}{%
  \bibinfo {author} {\bibfnamefont{Y.~F.}\ \bibnamefont{Duan}}\ and\ \bibinfo
  {author} {\bibfnamefont{K.~L.}\ \bibnamefont{Yao}},\ }%
  \bibfield{journal}{%
  \Doi{10.1103/PhysRevB.63.134434}{\bibinfo {journal} {Phys. Rev. B}}\ }%
  \textbf{\bibinfo {volume} {63}},\ \bibinfo {pages} {134434} (\bibinfo {month}
  {Mar}\ \bibinfo {year} {2001})%
  \bibAnnoteFile{NoStop}{Duan2001}%
\bibitem{Gul'acsi2005}%
  \BibitemOpen
  \bibfield{author}{%
  \bibinfo {author} {\bibfnamefont{Z.}~\bibnamefont{Gul\'acsi}}\ and\ \bibinfo
  {author} {\bibfnamefont{D.}~\bibnamefont{Vollhardt}},\ }%
  \bibfield{journal}{%
  \Doi{10.1103/PhysRevB.72.075130}{\bibinfo {journal} {Phys. Rev. B}}\ }%
  \textbf{\bibinfo {volume} {72}},\ \bibinfo {pages} {075130} (\bibinfo {month}
  {Aug}\ \bibinfo {year} {2005})%
  \bibAnnoteFile{NoStop}{Gul'acsi2005}%
\bibitem{Gul'acsi2003}%
  \BibitemOpen
  \bibfield{author}{%
  \bibinfo {author} {\bibfnamefont{Z.}~\bibnamefont{Gul\'acsi}}\ and\ \bibinfo
  {author} {\bibfnamefont{D.}~\bibnamefont{Vollhardt}},\ }%
  \bibfield{journal}{%
  \Doi{10.1103/PhysRevLett.91.186401}{\bibinfo {journal} {Phys. Rev. Lett.}}\
  }%
  \textbf{\bibinfo {volume} {91}},\ \bibinfo {pages} {186401} (\bibinfo {month}
  {Oct}\ \bibinfo {year} {2003})%
  \bibAnnoteFile{NoStop}{Gul'acsi2003}%
\bibitem{Richter2004a}%
  \BibitemOpen
  \bibfield{author}{%
  \bibinfo {author} {\bibfnamefont{J.}~\bibnamefont{Richter}}, \bibinfo
  {author} {\bibfnamefont{J.}~\bibnamefont{Schulenburg}}, \bibinfo {author}
  {\bibfnamefont{A.}~\bibnamefont{Honecker}}, \bibinfo {author}
  {\bibfnamefont{J.}~\bibnamefont{Schnack}},\ and\ \bibinfo {author}
  {\bibfnamefont{H.-J.}\ \bibnamefont{Schmidt}},\ }%
  \bibfield{journal}{%
  \Doi{10.1088/0953-8984/16/11/029}{\bibinfo {journal} {J. Phys.: Condens.
  Matter}}\ }%
  \textbf{\bibinfo {volume} {16}},\ \bibinfo {pages} {S779} (\bibinfo {year}
  {2004})%
  \bibAnnoteFile{NoStop}{Richter2004a}%
\bibitem{Richter2004}%
  \BibitemOpen
  \bibfield{author}{%
  \bibinfo {author} {\bibfnamefont{J.}~\bibnamefont{Richter}}, \bibinfo
  {author} {\bibfnamefont{O.}~\bibnamefont{Derzhko}},\ and\ \bibinfo {author}
  {\bibfnamefont{J.}~\bibnamefont{Schulenburg}},\ }%
  \bibfield{journal}{%
  \Doi{10.1103/PhysRevLett.93.107206}{\bibinfo {journal} {Phys. Rev. Lett.}}\
  }%
  \textbf{\bibinfo {volume} {93}},\ \bibinfo {pages} {107206} (\bibinfo {month}
  {Sep}\ \bibinfo {year} {2004})%
  \bibAnnoteFile{NoStop}{Richter2004}%
\bibitem{Rule2008}%
  \BibitemOpen
  \bibfield{author}{%
  \bibinfo {author} {\bibfnamefont{K.~C.}\ \bibnamefont{Rule}}, \bibinfo
  {author} {\bibfnamefont{A.~U.~B.}\ \bibnamefont{Wolter}}, \bibinfo {author}
  {\bibfnamefont{S.}~\bibnamefont{S\"ullow}}, \bibinfo {author}
  {\bibfnamefont{D.~A.}\ \bibnamefont{Tennant}}, \bibinfo {author}
  {\bibfnamefont{A.}~\bibnamefont{Br\"uhl}}, \bibinfo {author}
  {\bibfnamefont{S.}~\bibnamefont{K\"ohler}}, \bibinfo {author}
  {\bibfnamefont{B.}~\bibnamefont{Wolf}}, \bibinfo {author}
  {\bibfnamefont{M.}~\bibnamefont{Lang}},\ and\ \bibinfo {author}
  {\bibfnamefont{J.}~\bibnamefont{Schreuer}},\ }%
  \bibfield{journal}{%
  \Doi{10.1103/PhysRevLett.100.117202}{\bibinfo {journal} {Phys. Rev. Lett.}}\
  }%
  \textbf{\bibinfo {volume} {100}},\ \bibinfo {pages} {117202} (\bibinfo
  {month} {Mar}\ \bibinfo {year} {2008})%
  \bibAnnoteFile{NoStop}{Rule2008}%
\bibitem{Schulenburg2002}%
  \BibitemOpen
  \bibfield{author}{%
  \bibinfo {author} {\bibfnamefont{J.}~\bibnamefont{Schulenburg}}, \bibinfo
  {author} {\bibfnamefont{A.}~\bibnamefont{Honecker}}, \bibinfo {author}
  {\bibfnamefont{J.}~\bibnamefont{Schnack}}, \bibinfo {author}
  {\bibfnamefont{J.}~\bibnamefont{Richter}},\ and\ \bibinfo {author}
  {\bibfnamefont{H.-J.}\ \bibnamefont{Schmidt}},\ }%
  \bibfield{journal}{%
  \Doi{10.1103/PhysRevLett.88.167207}{\bibinfo {journal} {Phys. Rev. Lett.}}\
  }%
  \textbf{\bibinfo {volume} {88}},\ \bibinfo {pages} {167207} (\bibinfo {month}
  {Apr}\ \bibinfo {year} {2002})%
  \bibAnnoteFile{NoStop}{Schulenburg2002}%
\bibitem{Wu2007}%
  \BibitemOpen
  \bibfield{author}{%
  \bibinfo {author} {\bibfnamefont{C.}~\bibnamefont{Wu}}, \bibinfo {author}
  {\bibfnamefont{D.}~\bibnamefont{Bergman}}, \bibinfo {author}
  {\bibfnamefont{L.}~\bibnamefont{Balents}},\ and\ \bibinfo {author}
  {\bibfnamefont{S.}~\bibnamefont{Das~Sarma}},\ }%
  \bibfield{journal}{%
  \Doi{10.1103/PhysRevLett.99.070401}{\bibinfo {journal} {Phys. Rev. Lett.}}\
  }%
  \textbf{\bibinfo {volume} {99}},\ \bibinfo {pages} {070401} (\bibinfo {month}
  {Aug}\ \bibinfo {year} {2007})%
  \bibAnnoteFile{NoStop}{Wu2007}%
\bibitem{Wu2008}%
  \BibitemOpen
  \bibfield{author}{%
  \bibinfo {author} {\bibfnamefont{C.}~\bibnamefont{Wu}}\ and\ \bibinfo
  {author} {\bibfnamefont{S.}~\bibnamefont{Das~Sarma}},\ }%
  \bibfield{journal}{%
  \Doi{10.1103/PhysRevB.77.235107}{\bibinfo {journal} {Phys. Rev. B}}\ }%
  \textbf{\bibinfo {volume} {77}},\ \bibinfo {pages} {235107} (\bibinfo {month}
  {Jun}\ \bibinfo {year} {2008})%
  \bibAnnoteFile{NoStop}{Wu2008}%
\bibitem{Zhitomirsky2007}%
  \BibitemOpen
  \bibfield{author}{%
  \bibinfo {author} {\bibfnamefont{M.~E.}\ \bibnamefont{Zhitomirsky}}\ and\
  \bibinfo {author} {\bibfnamefont{H.}~\bibnamefont{Tsunetsugu}},\ }%
  \bibfield{journal}{%
  \Doi{10.1103/PhysRevB.75.224416}{\bibinfo {journal} {Phys. Rev. B}}\ }%
  \textbf{\bibinfo {volume} {75}},\ \bibinfo {pages} {224416} (\bibinfo {month}
  {Jun}\ \bibinfo {year} {2007})%
  \bibAnnoteFile{NoStop}{Zhitomirsky2007}%
\bibitem{Silvestre1985}%
  \BibitemOpen
  \bibfield{author}{%
  \bibinfo {author} {\bibfnamefont{J.}~\bibnamefont{Silvestre}}\ and\ \bibinfo
  {author} {\bibfnamefont{R.}~\bibnamefont{Hoffmann}},\ }%
  \bibfield{journal}{%
  \Doi{10.1021/ic00218a029}{\bibinfo {journal} {Inorg. Chem.}}\ }%
  \textbf{\bibinfo {volume} {24}},\ \bibinfo {pages} {4108} (\bibinfo {month}
  {Nov.}\ \bibinfo {year} {1985})%
  \bibAnnoteFile{NoStop}{Silvestre1985}%
\bibitem{Zheng2007}%
  \BibitemOpen
  \bibfield{author}{%
  \bibinfo {author} {\bibfnamefont{Y.-Z.}\ \bibnamefont{Zheng}}, \bibinfo
  {author} {\bibfnamefont{M.-L.}\ \bibnamefont{Tong}}, \bibinfo {author}
  {\bibfnamefont{W.}~\bibnamefont{Xue}}, \bibinfo {author}
  {\bibfnamefont{W.-X.}\ \bibnamefont{Zhang}}, \bibinfo {author}
  {\bibfnamefont{X.-M.}\ \bibnamefont{Chen}}, \bibinfo {author}
  {\bibfnamefont{F.}~\bibnamefont{Grandjean}},\ and\ \bibinfo {author}
  {\bibfnamefont{G.}~\bibnamefont{Long}},\ }%
  \bibfield{journal}{%
  \Doi{10.1002/anie.200701954}{\bibinfo {journal} {Angew. Chem. Int. Edit.}}\
  }%
  \textbf{\bibinfo {volume} {46}},\ \bibinfo {pages} {6076} (\bibinfo {year}
  {2007})%
  \bibAnnoteFile{NoStop}{Zheng2007}%
\bibitem{Jordan1928}%
  \BibitemOpen
  \bibfield{author}{%
  \bibinfo {author} {\bibfnamefont{P.}~\bibnamefont{Jordan}}\ and\ \bibinfo
  {author} {\bibfnamefont{E.}~\bibnamefont{Wigner}},\ }%
  \bibfield{journal}{%
  \Doi{10.1007/BF01331938}{\bibinfo {journal} {Z. Phys.}}\ }%
  \textbf{\bibinfo {volume} {47}},\ \bibinfo {pages} {631} (\bibinfo {year}
  {1928})%
  \bibAnnoteFile{NoStop}{Jordan1928}%
\bibitem{Baxter1973}%
  \BibitemOpen
  \bibfield{author}{%
  \bibinfo {author} {\bibfnamefont{R.}~\bibnamefont{Baxter}},\ }%
  \bibfield{journal}{%
  \Doi{DOI: 10.1016/0003-4916(73)90441-7}{\bibinfo {journal} {Ann. Phys.}}\ }%
  \textbf{\bibinfo {volume} {76}},\ \bibinfo {pages} {48 } (\bibinfo {year}
  {1973})%
  \bibAnnoteFile{NoStop}{Baxter1973}%
\bibitem{Baxter1973a}%
  \BibitemOpen
  \bibfield{author}{%
  \bibinfo {author} {\bibfnamefont{R.}~\bibnamefont{Baxter}},\ }%
  \bibfield{journal}{%
  \Doi{DOI: 10.1016/0003-4916(73)90440-5}{\bibinfo {journal} {Ann. Phys.}}\ }%
  \textbf{\bibinfo {volume} {76}},\ \bibinfo {pages} {25 } (\bibinfo {year}
  {1973})%
  \bibAnnoteFile{NoStop}{Baxter1973a}%
\bibitem{Baxter1973b}%
  \BibitemOpen
  \bibfield{author}{%
  \bibinfo {author} {\bibfnamefont{R.}~\bibnamefont{Baxter}},\ }%
  \bibfield{journal}{%
  \Doi{DOI: 10.1016/0003-4916(73)90439-9}{\bibinfo {journal} {Ann. Phys.}}\ }%
  \textbf{\bibinfo {volume} {76}},\ \bibinfo {pages} {1 } (\bibinfo {year}
  {1973})%
  \bibAnnoteFile{NoStop}{Baxter1973b}%
\bibitem{Bethe1931}%
  \BibitemOpen
  \bibfield{author}{%
  \bibinfo {author} {\bibfnamefont{H.}~\bibnamefont{Bethe}},\ }%
  \bibfield{journal}{%
  \Doi{10.1007/BF01341708}{\bibinfo {journal} {Z. Phys.}}\ }%
  \textbf{\bibinfo {volume} {71}},\ \bibinfo {pages} {205} (\bibinfo {year}
  {1931})%
  \bibAnnoteFile{NoStop}{Bethe1931}%
\bibitem{Fradkin1989}%
  \BibitemOpen
  \bibfield{author}{%
  \bibinfo {author} {\bibfnamefont{E.}~\bibnamefont{Fradkin}},\ }%
  \bibfield{journal}{%
  \Doi{10.1103/PhysRevLett.63.322}{\bibinfo {journal} {Phys. Rev. Lett.}}\ }%
  \textbf{\bibinfo {volume} {63}},\ \bibinfo {pages} {322} (\bibinfo {month}
  {Jul}\ \bibinfo {year} {1989})%
  \bibAnnoteFile{NoStop}{Fradkin1989}%
\bibitem{Wang1992}%
  \BibitemOpen
  \bibfield{author}{%
  \bibinfo {author} {\bibfnamefont{Y.~R.}\ \bibnamefont{Wang}},\ }%
  \bibfield{journal}{%
  \Doi{10.1103/PhysRevB.45.12604}{\bibinfo {journal} {Phys. Rev. B}}\ }%
  \textbf{\bibinfo {volume} {45}},\ \bibinfo {pages} {12604} (\bibinfo {month}
  {Jun}\ \bibinfo {year} {1992})%
  \bibAnnoteFile{NoStop}{Wang1992}%
\bibitem{Wang1992a}%
  \BibitemOpen
  \bibfield{author}{%
  \bibinfo {author} {\bibfnamefont{Y.~R.}\ \bibnamefont{Wang}},\ }%
  \bibfield{journal}{%
  \Doi{10.1103/PhysRevB.46.151}{\bibinfo {journal} {Phys. Rev. B}}\ }%
  \textbf{\bibinfo {volume} {46}},\ \bibinfo {pages} {151} (\bibinfo {month}
  {Jul}\ \bibinfo {year} {1992})%
  \bibAnnoteFile{NoStop}{Wang1992a}%
\bibitem{Wang1991}%
  \BibitemOpen
  \bibfield{author}{%
  \bibinfo {author} {\bibfnamefont{Y.~R.}\ \bibnamefont{Wang}},\ }%
  \bibfield{journal}{%
  \Doi{10.1103/PhysRevB.43.3786}{\bibinfo {journal} {Phys. Rev. B}}\ }%
  \textbf{\bibinfo {volume} {43}},\ \bibinfo {pages} {3786} (\bibinfo {month}
  {Feb}\ \bibinfo {year} {1991})%
  \bibAnnoteFile{NoStop}{Wang1991}%
\bibitem{Ambjoern1989}%
  \BibitemOpen
  \bibfield{author}{%
  \bibinfo {author} {\bibfnamefont{J.}~\bibnamefont{Ambjørn}}\ and\ \bibinfo
  {author} {\bibfnamefont{G.~W.}\ \bibnamefont{Semenoff}},\ }%
  \bibfield{journal}{%
  \Doi{DOI: 10.1016/0370-2693(89)90296-7}{\bibinfo {journal} {Phys. Lett. B}}\
  }%
  \textbf{\bibinfo {volume} {226}},\ \bibinfo {pages} {107 } (\bibinfo {year}
  {1989})%
  \bibAnnoteFile{NoStop}{Ambjoern1989}%
\bibitem{Moessner2006}%
  \BibitemOpen
  \bibfield{author}{%
  \bibinfo {author} {\bibfnamefont{R.}~\bibnamefont{Moessner}}\ and\ \bibinfo
  {author} {\bibfnamefont{A.~P.}\ \bibnamefont{Ramirez}},\ }%
  \bibfield{journal}{%
  \Doi{10.1063/1.2186278}{\bibinfo {journal} {Phys. Today}}\ }%
  \textbf{\bibinfo {volume} {59}},\ \bibinfo {pages} {24} (\bibinfo {year}
  {2006})%
  \bibAnnoteFile{NoStop}{Moessner2006}%
\bibitem{Haldane1980}%
  \BibitemOpen
  \bibfield{author}{%
  \bibinfo {author} {\bibfnamefont{F.~D.~M.}\ \bibnamefont{Haldane}},\ }%
  \bibfield{journal}{%
  \Doi{10.1103/PhysRevLett.45.1358}{\bibinfo {journal} {Phys. Rev. Lett.}}\ }%
  \textbf{\bibinfo {volume} {45}},\ \bibinfo {pages} {1358} (\bibinfo {month}
  {Oct}\ \bibinfo {year} {1980})%
  \bibAnnoteFile{NoStop}{Haldane1980}%
\bibitem{Nakamura1997}%
  \BibitemOpen
  \bibfield{author}{%
  \bibinfo {author} {\bibfnamefont{M.}~\bibnamefont{Nakamura}}\ and\ \bibinfo
  {author} {\bibfnamefont{K.}~\bibnamefont{Nomura}},\ }%
  \bibfield{journal}{%
  \Doi{10.1103/PhysRevB.56.12840}{\bibinfo {journal} {Phys. Rev. B}}\ }%
  \textbf{\bibinfo {volume} {56}},\ \bibinfo {pages} {12840} (\bibinfo {month}
  {Nov}\ \bibinfo {year} {1997})%
  \bibAnnoteFile{NoStop}{Nakamura1997}%
\bibitem{Doucot2002}%
  \BibitemOpen
  \bibfield{author}{%
  \bibinfo {author}
  {\bibfnamefont{B.}~\bibnamefont{Dou\ifmmode~\mbox{\c{c}}\else \c{c}\fi{}ot}}\
  and\ \bibinfo {author} {\bibfnamefont{J.}~\bibnamefont{Vidal}},\ }%
  \bibfield{journal}{%
  \Doi{10.1103/PhysRevLett.88.227005}{\bibinfo {journal} {Phys. Rev. Lett.}}\
  }%
  \textbf{\bibinfo {volume} {88}},\ \bibinfo {pages} {227005} (\bibinfo {month}
  {May}\ \bibinfo {year} {2002})%
  \bibAnnoteFile{NoStop}{Doucot2002}%
\bibitem{Gulacsi2008}%
  \BibitemOpen
  \bibfield{author}{%
  \bibinfo {author} {\bibfnamefont{Z.}~\bibnamefont{Gul\'acsi}}, \bibinfo
  {author} {\bibfnamefont{A.}~\bibnamefont{Kampf}},\ and\ \bibinfo {author}
  {\bibfnamefont{D.}~\bibnamefont{Vollhardt}},\ }%
  \bibfield{journal}{%
  \Doi{10.1143/PTPS.176.1}{\bibinfo {journal} {Prog. Theor. Phys. Supp.}}\ }%
  \textbf{\bibinfo {volume} {176}},\ \bibinfo {pages} {1} (\bibinfo {year}
  {2008})%
  \bibAnnoteFile{NoStop}{Gulacsi2008}%
\bibitem{Aoki1996}%
  \BibitemOpen
  \bibfield{author}{%
  \bibinfo {author} {\bibfnamefont{H.}~\bibnamefont{Aoki}}, \bibinfo {author}
  {\bibfnamefont{M.}~\bibnamefont{Ando}},\ and\ \bibinfo {author}
  {\bibfnamefont{H.}~\bibnamefont{Matsumura}},\ }%
  \bibfield{journal}{%
  \Doi{10.1103/PhysRevB.54.R17296}{\bibinfo {journal} {Phys. Rev. B}}\ }%
  \textbf{\bibinfo {volume} {54}},\ \bibinfo {pages} {R17296} (\bibinfo {month}
  {Dec}\ \bibinfo {year} {1996})%
  \bibAnnoteFile{NoStop}{Aoki1996}%
\bibitem{Lieb1989}%
  \BibitemOpen
  \bibfield{author}{%
  \bibinfo {author} {\bibfnamefont{E.~H.}\ \bibnamefont{Lieb}},\ }%
  \bibfield{journal}{%
  \Doi{10.1103/PhysRevLett.62.1201}{\bibinfo {journal} {Phys. Rev. Lett.}}\ }%
  \textbf{\bibinfo {volume} {62}},\ \bibinfo {pages} {1201} (\bibinfo {month}
  {Mar}\ \bibinfo {year} {1989})%
  \bibAnnoteFile{NoStop}{Lieb1989}%
\bibitem{Dias2000}%
  \BibitemOpen
  \bibfield{author}{%
  \bibinfo {author} {\bibfnamefont{R.~G.}\ \bibnamefont{Dias}},\ }%
  \bibfield{journal}{%
  \Doi{10.1103/PhysRevB.62.7791}{\bibinfo {journal} {Phys. Rev. B}}\ }%
  \textbf{\bibinfo {volume} {62}},\ \bibinfo {pages} {7791} (\bibinfo {month}
  {Sep}\ \bibinfo {year} {2000})%
  \bibAnnoteFile{NoStop}{Dias2000}%
\bibitem{G'omez-Santos1993}%
  \BibitemOpen
  \bibfield{author}{%
  \bibinfo {author} {\bibfnamefont{G.}~\bibnamefont{G\'omez-Santos}},\ }%
  \bibfield{journal}{%
  \Doi{10.1103/PhysRevLett.70.3780}{\bibinfo {journal} {Phys. Rev. Lett.}}\ }%
  \textbf{\bibinfo {volume} {70}},\ \bibinfo {pages} {3780} (\bibinfo {month}
  {Jun}\ \bibinfo {year} {1993})%
  \bibAnnoteFile{NoStop}{G'omez-Santos1993}%
\bibitem{G'omez-Santos1992}%
  \BibitemOpen
  \bibfield{author}{%
  \bibinfo {author} {\bibfnamefont{G.}~\bibnamefont{G\'omez-Santos}},\ }%
  \bibfield{journal}{%
  \Doi{10.1103/PhysRevB.46.14217}{\bibinfo {journal} {Phys. Rev. B}}\ }%
  \textbf{\bibinfo {volume} {46}},\ \bibinfo {pages} {14217} (\bibinfo {month}
  {Dec}\ \bibinfo {year} {1992})%
  \bibAnnoteFile{NoStop}{G'omez-Santos1992}%
\bibitem{Emery1990}%
  \BibitemOpen
  \bibfield{author}{%
  \bibinfo {author} {\bibfnamefont{V.~J.}\ \bibnamefont{Emery}}, \bibinfo
  {author} {\bibfnamefont{S.~A.}\ \bibnamefont{Kivelson}},\ and\ \bibinfo
  {author} {\bibfnamefont{H.~Q.}\ \bibnamefont{Lin}},\ }%
  \bibfield{journal}{%
  \Doi{10.1103/PhysRevLett.64.475}{\bibinfo {journal} {Phys. Rev. Lett.}}\ }%
  \textbf{\bibinfo {volume} {64}},\ \bibinfo {pages} {475} (\bibinfo {month}
  {Jan}\ \bibinfo {year} {1990})%
  \bibAnnoteFile{NoStop}{Emery1990}%
\bibitem{Ogata1991}%
  \BibitemOpen
  \bibfield{author}{%
  \bibinfo {author} {\bibfnamefont{M.}~\bibnamefont{Ogata}}, \bibinfo {author}
  {\bibfnamefont{M.~U.}\ \bibnamefont{Luchini}}, \bibinfo {author}
  {\bibfnamefont{S.}~\bibnamefont{Sorella}},\ and\ \bibinfo {author}
  {\bibfnamefont{F.~F.}\ \bibnamefont{Assaad}},\ }%
  \bibfield{journal}{%
  \Doi{10.1103/PhysRevLett.66.2388}{\bibinfo {journal} {Phys. Rev. Lett.}}\ }%
  \textbf{\bibinfo {volume} {66}},\ \bibinfo {pages} {2388} (\bibinfo {month}
  {May}\ \bibinfo {year} {1991})%
  \bibAnnoteFile{NoStop}{Ogata1991}%
\bibitem{Haldane1981}%
  \BibitemOpen
  \bibfield{author}{%
  \bibinfo {author} {\bibfnamefont{F.~D.~M.}\ \bibnamefont{Haldane}},\ }%
  \bibfield{journal}{%
  \Doi{10.1103/PhysRevLett.47.1840}{\bibinfo {journal} {Phys. Rev. Lett.}}\ }%
  \textbf{\bibinfo {volume} {47}},\ \bibinfo {pages} {1840} (\bibinfo {month}
  {Dec}\ \bibinfo {year} {1981})%
  \bibAnnoteFile{NoStop}{Haldane1981}%
\bibitem{Haldane1981a}%
  \BibitemOpen
  \bibfield{author}{%
  \bibinfo {author} {\bibfnamefont{F.~D.~M.}\ \bibnamefont{Haldane}},\ }%
  \bibfield{journal}{%
  \bibinfo {journal} {J. Phys. C: Solid State}\ }%
  \textbf{\bibinfo {volume} {14}},\ \bibinfo {pages} {2585} (\bibinfo {year}
  {1981})%
  \bibAnnoteFile{NoStop}{Haldane1981a}%
\bibitem{Fabrizio1995}%
  \BibitemOpen
  \bibfield{author}{%
  \bibinfo {author} {\bibfnamefont{M.}~\bibnamefont{Fabrizio}}\ and\ \bibinfo
  {author} {\bibfnamefont{A.~O.}\ \bibnamefont{Gogolin}},\ }%
  \bibfield{journal}{%
  \Doi{10.1103/PhysRevB.51.17827}{\bibinfo {journal} {Phys. Rev. B}}\ }%
  \textbf{\bibinfo {volume} {51}},\ \bibinfo {pages} {17827} (\bibinfo {month}
  {Jun}\ \bibinfo {year} {1995})%
  \bibAnnoteFile{NoStop}{Fabrizio1995}%
\bibitem{Eggert1996}%
  \BibitemOpen
  \bibfield{author}{%
  \bibinfo {author} {\bibfnamefont{S.}~\bibnamefont{Eggert}}, \bibinfo {author}
  {\bibfnamefont{H.}~\bibnamefont{Johannesson}},\ and\ \bibinfo {author}
  {\bibfnamefont{A.}~\bibnamefont{Mattsson}},\ }%
  \bibfield{journal}{%
  \Doi{10.1103/PhysRevLett.76.1505}{\bibinfo {journal} {Phys. Rev. Lett.}}\ }%
  \textbf{\bibinfo {volume} {76}},\ \bibinfo {pages} {1505} (\bibinfo {month}
  {Feb}\ \bibinfo {year} {1996})%
  \bibAnnoteFile{NoStop}{Eggert1996}%
\bibitem{Wang1996}%
  \BibitemOpen
  \bibfield{author}{%
  \bibinfo {author} {\bibfnamefont{Y.}~\bibnamefont{Wang}}, \bibinfo {author}
  {\bibfnamefont{J.}~\bibnamefont{Voit}},\ and\ \bibinfo {author}
  {\bibfnamefont{F.-C.}\ \bibnamefont{Pu}},\ }%
  \bibfield{journal}{%
  \Doi{10.1103/PhysRevB.54.8491}{\bibinfo {journal} {Phys. Rev. B}}\ }%
  \textbf{\bibinfo {volume} {54}},\ \bibinfo {pages} {8491} (\bibinfo {month}
  {Sep}\ \bibinfo {year} {1996})%
  \bibAnnoteFile{NoStop}{Wang1996}%
\bibitem{Voit2000}%
  \BibitemOpen
  \bibfield{author}{%
  \bibinfo {author} {\bibfnamefont{J.}~\bibnamefont{Voit}}, \bibinfo {author}
  {\bibfnamefont{Y.}~\bibnamefont{Wang}},\ and\ \bibinfo {author}
  {\bibfnamefont{M.}~\bibnamefont{Grioni}},\ }%
  \bibfield{journal}{%
  \Doi{10.1103/PhysRevB.61.7930}{\bibinfo {journal} {Phys. Rev. B}}\ }%
  \textbf{\bibinfo {volume} {61}},\ \bibinfo {pages} {7930} (\bibinfo {month}
  {Mar}\ \bibinfo {year} {2000})%
  \bibAnnoteFile{NoStop}{Voit2000}%
\bibitem{Meden2000}%
  \BibitemOpen
  \bibfield{author}{%
  \bibinfo {author} {\bibfnamefont{V.}~\bibnamefont{Meden}}, \bibinfo {author}
  {\bibfnamefont{W.}~\bibnamefont{Metzner}}, \bibinfo {author}
  {\bibfnamefont{U.}~\bibnamefont{Schollwöck}}, \bibinfo {author}
  {\bibfnamefont{O.}~\bibnamefont{Schneider}}, \bibinfo {author}
  {\bibfnamefont{T.}~\bibnamefont{Stauber}},\ and\ \bibinfo {author}
  {\bibfnamefont{K.}~\bibnamefont{Schönhammer}},\ }%
  \bibfield{journal}{%
  \Doi{10.1007/s100510070180}{\bibinfo {journal} {Europ. Phys. J. B}}\ }%
  \textbf{\bibinfo {volume} {16}},\ \bibinfo {pages} {631} (\bibinfo {year}
  {2000})%
  \bibAnnoteFile{NoStop}{Meden2000}%
\bibitem{Kane1992}%
  \BibitemOpen
  \bibfield{author}{%
  \bibinfo {author} {\bibfnamefont{C.~L.}\ \bibnamefont{Kane}}\ and\ \bibinfo
  {author} {\bibfnamefont{M.~P.~A.}\ \bibnamefont{Fisher}},\ }%
  \bibfield{journal}{%
  \Doi{10.1103/PhysRevLett.68.1220}{\bibinfo {journal} {Phys. Rev. Lett.}}\ }%
  \textbf{\bibinfo {volume} {68}},\ \bibinfo {pages} {1220} (\bibinfo {month}
  {Feb}\ \bibinfo {year} {1992})%
  \bibAnnoteFile{NoStop}{Kane1992}%
\bibitem{Kane1992a}%
  \BibitemOpen
  \bibfield{author}{%
  \bibinfo {author} {\bibfnamefont{C.~L.}\ \bibnamefont{Kane}}\ and\ \bibinfo
  {author} {\bibfnamefont{M.~P.~A.}\ \bibnamefont{Fisher}},\ }%
  \bibfield{journal}{%
  \Doi{10.1103/PhysRevB.46.15233}{\bibinfo {journal} {Phys. Rev. B}}\ }%
  \textbf{\bibinfo {volume} {46}},\ \bibinfo {pages} {15233} (\bibinfo {month}
  {Dec}\ \bibinfo {year} {1992})%
  \bibAnnoteFile{NoStop}{Kane1992a}%
\bibitem{Kohn1964}%
  \BibitemOpen
  \bibfield{author}{%
  \bibinfo {author} {\bibfnamefont{W.}~\bibnamefont{Kohn}},\ }%
  \bibfield{journal}{%
  \Doi{10.1103/PhysRev.133.A171}{\bibinfo {journal} {Phys. Rev.}}\ }%
  \textbf{\bibinfo {volume} {133}},\ \bibinfo {pages} {A171} (\bibinfo {month}
  {Jan}\ \bibinfo {year} {1964})%
  \bibAnnoteFile{NoStop}{Kohn1964}%
\bibitem{Shastry1990}%
  \BibitemOpen
  \bibfield{author}{%
  \bibinfo {author} {\bibfnamefont{B.~S.}\ \bibnamefont{Shastry}}\ and\
  \bibinfo {author} {\bibfnamefont{B.}~\bibnamefont{Sutherland}},\ }%
  \bibfield{journal}{%
  \Doi{10.1103/PhysRevLett.65.243}{\bibinfo {journal} {Phys. Rev. Lett.}}\ }%
  \textbf{\bibinfo {volume} {65}},\ \bibinfo {pages} {243} (\bibinfo {month}
  {Jul}\ \bibinfo {year} {1990})%
  \bibAnnoteFile{NoStop}{Shastry1990}%
\bibitem{Fye1991}%
  \BibitemOpen
  \bibfield{author}{%
  \bibinfo {author} {\bibfnamefont{R.~M.}\ \bibnamefont{Fye}}, \bibinfo
  {author} {\bibfnamefont{M.~J.}\ \bibnamefont{Martins}}, \bibinfo {author}
  {\bibfnamefont{D.~J.}\ \bibnamefont{Scalapino}}, \bibinfo {author}
  {\bibfnamefont{J.}~\bibnamefont{Wagner}},\ and\ \bibinfo {author}
  {\bibfnamefont{W.}~\bibnamefont{Hanke}},\ }%
  \bibfield{journal}{%
  \Doi{10.1103/PhysRevB.44.6909}{\bibinfo {journal} {Phys. Rev. B}}\ }%
  \textbf{\bibinfo {volume} {44}},\ \bibinfo {pages} {6909} (\bibinfo {month}
  {Oct}\ \bibinfo {year} {1991})%
  \bibAnnoteFile{NoStop}{Fye1991}%
\bibitem{Ogata1990}%
  \BibitemOpen
  \bibfield{author}{%
  \bibinfo {author} {\bibfnamefont{M.}~\bibnamefont{Ogata}}\ and\ \bibinfo
  {author} {\bibfnamefont{H.}~\bibnamefont{Shiba}},\ }%
  \bibfield{journal}{%
  \Doi{10.1103/PhysRevB.41.2326}{\bibinfo {journal} {Phys. Rev. B}}\ }%
  \textbf{\bibinfo {volume} {41}},\ \bibinfo {pages} {2326} (\bibinfo {month}
  {Feb}\ \bibinfo {year} {1990})%
  \bibAnnoteFile{NoStop}{Ogata1990}%
\bibitem{Schadschneider1995}%
  \BibitemOpen
  \bibfield{author}{%
  \bibinfo {author} {\bibfnamefont{A.}~\bibnamefont{Schadschneider}},\ }%
  \bibfield{journal}{%
  \Doi{10.1103/PhysRevB.51.10386}{\bibinfo {journal} {Phys. Rev. B}}\ }%
  \textbf{\bibinfo {volume} {51}},\ \bibinfo {pages} {10386} (\bibinfo {month}
  {Apr}\ \bibinfo {year} {1995})%
  \bibAnnoteFile{NoStop}{Schadschneider1995}%
\bibitem{Gebhard1997}%
  \BibitemOpen
  \bibfield{author}{%
  \bibinfo {author} {\bibfnamefont{F.}~\bibnamefont{Gebhard}}, \bibinfo
  {author} {\bibfnamefont{K.}~\bibnamefont{Born}}, \bibinfo {author}
  {\bibfnamefont{M.}~\bibnamefont{Scheidler}}, \bibinfo {author}
  {\bibfnamefont{P.}~\bibnamefont{Thomas}},\ and\ \bibinfo {author}
  {\bibfnamefont{S.~W.}\ \bibnamefont{Koch}},\ }%
  \bibfield{journal}{%
  \Doi{10.1080/13642819708205701}{\bibinfo {journal} {Phil. Mag. B}}\ }%
  \textbf{\bibinfo {volume} {75}},\ \bibinfo {pages} {13} (\bibinfo {year}
  {1997}),\ ISSN \bibinfo {issn} {1364-2812}%
  \bibAnnoteFile{NoStop}{Gebhard1997}%
\bibitem{Dias1992}%
  \BibitemOpen
  \bibfield{author}{%
  \bibinfo {author} {\bibnamefont{{Ricardo G. Dias}}}\ and\ \bibinfo {author}
  {\bibnamefont{{J. M. B. Lopes dos Santos}}},\ }%
  \bibfield{journal}{%
  \Doi{10.1051/jp1:1992252}{\bibinfo {journal} {J. Phys. I France}}\ }%
  \textbf{\bibinfo {volume} {2}},\ \bibinfo {pages} {1889} (\bibinfo {year}
  {1992})%
  \bibAnnoteFile{NoStop}{Dias1992}%
\bibitem{Peres2000}%
  \BibitemOpen
  \bibfield{author}{%
  \bibinfo {author} {\bibfnamefont{N.~M.~R.}\ \bibnamefont{Peres}}, \bibinfo
  {author} {\bibfnamefont{R.~G.}\ \bibnamefont{Dias}}, \bibinfo {author}
  {\bibfnamefont{P.~D.}\ \bibnamefont{Sacramento}},\ and\ \bibinfo {author}
  {\bibfnamefont{J.~M.~P.}\ \bibnamefont{Carmelo}},\ }%
  \bibfield{journal}{%
  \Doi{10.1103/PhysRevB.61.5169}{\bibinfo {journal} {Phys. Rev. B}}\ }%
  \textbf{\bibinfo {volume} {61}},\ \bibinfo {pages} {5169} (\bibinfo {month}
  {Feb}\ \bibinfo {year} {2000})%
  \bibAnnoteFile{NoStop}{Peres2000}%
\bibitem{Arrachea1994}%
  \BibitemOpen
  \bibfield{author}{%
  \bibinfo {author} {\bibfnamefont{L.}~\bibnamefont{Arrachea}}\ and\ \bibinfo
  {author} {\bibfnamefont{A.~A.}\ \bibnamefont{Aligia}},\ }%
  \bibfield{journal}{%
  \Doi{10.1103/PhysRevLett.73.2240}{\bibinfo {journal} {Phys. Rev. Lett.}}\ }%
  \textbf{\bibinfo {volume} {73}},\ \bibinfo {pages} {2240} (\bibinfo {month}
  {Oct}\ \bibinfo {year} {1994})%
  \bibAnnoteFile{NoStop}{Arrachea1994}%
\bibitem{Harris1967}%
  \BibitemOpen
  \bibfield{author}{%
  \bibinfo {author} {\bibfnamefont{A.~B.}\ \bibnamefont{Harris}}\ and\ \bibinfo
  {author} {\bibfnamefont{R.~V.}\ \bibnamefont{Lange}},\ }%
  \bibfield{journal}{%
  \Doi{10.1103/PhysRev.157.295}{\bibinfo {journal} {Phys. Rev.}}\ }%
  \textbf{\bibinfo {volume} {157}},\ \bibinfo {pages} {295} (\bibinfo {month}
  {May}\ \bibinfo {year} {1967})%
  \bibAnnoteFile{NoStop}{Harris1967}%
\bibitem{Derzhko2001}%
  \BibitemOpen
  \bibfield{author}{%
  \bibinfo {author} {\bibfnamefont{O.}~\bibnamefont{Derzhko}},\ }%
  \bibfield{journal}{%
  \bibinfo {journal} {J. Phys. Stud. (L'viv)}\ }%
  \textbf{\bibinfo {volume} {5}},\ \bibinfo {pages} {49} (\bibinfo {year}
  {2001})%
  \bibAnnoteFile{NoStop}{Derzhko2001}%
\bibitem{Benoit2002}%
  \BibitemOpen
  \bibfield{author}{%
  \bibinfo {author}
  {\bibfnamefont{B.}~\bibnamefont{Dou\ifmmode~\mbox{\c{c}}\else \c{c}\fi{}ot}}\
  and\ \bibinfo {author} {\bibfnamefont{J.}~\bibnamefont{Vidal}},\ }%
  \bibfield{journal}{%
  \Doi{10.1103/PhysRevLett.88.227005}{\bibinfo {journal} {Phys. Rev. Lett.}}\
  }%
  \textbf{\bibinfo {volume} {88}},\ \bibinfo {pages} {227005} (\bibinfo {month}
  {May}\ \bibinfo {year} {2002})%
  \bibAnnoteFile{NoStop}{Benoit2002}%
\bibitem{Choi1985}%
  \BibitemOpen
  \bibfield{author}{%
  \bibinfo {author} {\bibfnamefont{M.~Y.}\ \bibnamefont{Choi}}\ and\ \bibinfo
  {author} {\bibfnamefont{S.}~\bibnamefont{Doniach}},\ }%
  \bibfield{journal}{%
  \Doi{10.1103/PhysRevB.31.4516}{\bibinfo {journal} {Phys. Rev. B}}\ }%
  \textbf{\bibinfo {volume} {31}},\ \bibinfo {pages} {4516} (\bibinfo {month}
  {Apr}\ \bibinfo {year} {1985})%
  \bibAnnoteFile{NoStop}{Choi1985}%
\end{thebibliography}%

\end{document}